\documentclass{iopjournal}
\usepackage{url}
\usepackage{lineno}
\usepackage{multirow}
\usepackage{booktabs}
\usepackage{amsfonts}
\usepackage{amsmath}
\usepackage{tabularx}

\newcommand{\arcond}{X^{t-2:t-1}}
\newcommand{\vardist}{q\left(Z^t \middle| X^{t-2:t-1}, X^{t} \right)}
\newcommand{\priordist}{p\left(Z^t \middle| \arcond \right)}

\newcommand{\varloss}{\mathcal{L}_\mathrm{Var}}
\newcommand{\crpsloss}{\mathcal{L}_\mathrm{CRPS}}
\newcommand{\divloss}{\mathcal{L}_\mathrm{Div}}

\begin{document}

\articletype{}

\title{Deterministic and probabilistic neural surrogates of\\global hybrid-Vlasov simulations}

\author{Daniel Holmberg$^{1,3,*}$\orcid{0000-0001-5020-7438}, Ivan Zaitsev$^2$\orcid{0000-0002-0640-0123}, Markku Alho$^2$, Ioanna Bouri$^1$,\\Fanni Franssila$^1$, Haewon Jeong$^3$, Minna Palmroth$^{2,4}$ and Teemu Roos$^1$}

\affil{$^1$Department of Computer Science, University of Helsinki}

\affil{$^2$Department of Physics, University of Helsinki}

\affil{$^3$Department of Electrical and Computer Engineering, University of California, Santa Barbara}

\affil{$^4$Space and Earth Observation Centre, Finnish Meteorological Institute}

\affil{$^*$Author to whom any correspondence should be addressed.}

\email{dholmberg@ucsb.edu}


\begin{abstract}
Hybrid-Vlasov simulations resolve ion-kinetic effects in the solar wind–magnetosphere interaction, but even 5D (2D + 3V) configurations are computationally expensive. We show that graph-based machine learning emulators can learn the spatiotemporal evolution of electromagnetic fields and lower-order moments of the ion velocity distribution function in near-Earth space from four 5D Vlasiator runs, each driven by steady solar wind conditions. The upstream ion number density is systematically varied between the runs, while the grid spacing is held constant, to scan the ratio of ion inertial length to grid size. Using a graph neural network (GNN) operating on the 2D spatial simulation grid comprising 670k cells, we demonstrate that both a deterministic forecasting model (Graph-FM) and a probabilistic ensemble forecasting model (Graph-EFM) based on a latent variable formulation produce accurate predictions of future plasma states. A divergence penalty is incorporated to encourage divergence-freeness in the magnetic fields. For the probabilistic model, a continuous ranked probability score objective is added to improve the calibration of the ensemble forecasts. In terms of wall time per output step, the trained emulators run over two orders of magnitude faster on a single GPU than the Vlasiator simulations on 100 CPUs, although this comparison involves different hardware and the emulators predict only moments of the ion velocity distribution function rather than the full distribution. Most forecasted fields have Pearson correlations above 0.95 at 50 seconds lead time. Fields that exhibit degenerate (near-zero) distributions in the 5D setting are more challenging for the emulator to keep well correlated. The ensemble forecasts remain underdispersive, with spread--skill ratios of approximately 0.2--0.3, and thus provide spatially structured relative uncertainty estimates. Overall, these results demonstrate that GNNs provide a viable framework for rapid ensemble generation in hybrid-Vlasov modeling and highlight promising directions for future work.
\end{abstract}

\section{Introduction}

The interaction between the solar wind and Earth’s magnetosphere represents a fundamental multi-scale plasma physics problem, characterized by the coupling of global magnetospheric dynamics with micro-scale ion-kinetic processes. While the large-scale evolution of the system is often described by magnetohydrodynamic (MHD) fluid approximations~\cite{janhunen2012gumics, glocer2013coupledbatsrus}, these models cannot resolve kinetic processes that govern energy dissipation and transport at plasma boundaries.

A range of kinetic and hybrid-kinetic simulation approaches, including hybrid particle-in-cell codes, are actively used to study global and local space plasmas in both 2D and 3D; see~\cite{shay2025simulation} for a recent overview of the state of the art in magnetospheric and reconnection modeling. Among these, hybrid-Vlasov models such as Vlasiator~\cite{vonalfthan2014vlasiator, palmroth2018vlasov, palmroth2025vlasov} investigate ion kinetics in Earth's magnetosphere by solving the Vlasov equation for ions on a phase-space grid while treating electrons as a charge-neutralizing fluid. This approach has enabled studies of complex plasma phenomena, including the transmission of foreshock waves through the bow shock~\cite{turc2023transmission}, interaction of magnetotail reconnection with cross-tail instabilities~\cite{palmroth2023magnetotail, cozzani2025interplay}, and the formation of magnetosheath jets~\cite{suni2025magnetosheath}. However, the high fidelity of these simulations comes at an immense computational cost. The massive resources required for even a single global Vlasiator run effectively prohibit high-throughput parameter studies, which are necessary for operational forecasting.

In an attempt to speed up numerical simulation, neural surrogates have emerged in recent years, with approaches based on neural operators that learn mappings between function spaces~\cite{li2020fourier, lu2021learning, kovachki2023neural}, convolutional neural networks with learned corrections embedded within differentiable solvers~\cite{kochkov2021machine}, or graph neural networks (GNNs) applicable to versatile geometries~\cite{pfaff2021learning, lino2021simulating}. Such approaches have been applied also for plasmas across a range of fidelities and geometries. Neural operators have been proposed as surrogates for MHD simulations~\cite{gopakumar2024plasma, carey2025neural}. GNNs have been applied to learn the dynamics of a one-dimensional plasma sheet model~\cite{carvalho2024learning}, and to emulate particle-in-cell simulations~\cite{mlinarevic2025particle}. GNNs have further been used to approximate elliptic subproblems such as the Poisson equation on unstructured grids for Hall thruster modeling~\cite{vigot2025graph}. In terms of space plasma, echo state networks have been used to emulate the auroral current system~\cite{kataoka2024machine}, GNNs have been developed to forecast total electron content in the ionosphere \cite{kelebek2025ioncast}, and the spherical Fourier neural operator has been adopted to predict the radial velocity of the solar wind in the heliosphere~\cite{mansouri2025toward}.

A limitation of these plasma surrogates is their reliance on deterministic single-point forecasts. The predictive fidelity of kinetic plasma simulations is significantly affected by uncertainties in model parameters, initial conditions, and external forcing, making uncertainty quantification an essential component of global plasma modeling~\cite{chen2025micro}. In other related fields that rely on MHD for making predictions, the need for uncertainty-aware modeling has been strongly advocated. One example is space weather forecasting, where it can enhance reliability and support more informed decision-making~\cite{murray2018importance}. Similarly, uncertainty quantification has been identified as an important direction for the development of surrogate models of MHD simulations in tokamak fusion research~\cite{gopakumar2024plasma}. Traditional ensemble methods for numerical simulation, which require perturbing initial conditions and performing multiple runs~\cite{morley2018perturbed}, are computationally prohibitive, and especially so for global kinetic solvers. Generative machine learning frameworks offer a path toward rapid, uncertainty-aware emulation of such systems. They also add their own uncertainty, from limited training data, an incomplete state representation, and finite model capacity.

To help tackle these challenges, we take inspiration from recent breakthroughs in machine learning for atmospheric weather forecasting~\cite{ben2024rise}, where both deterministic~\cite{keisler2022forecasting, bi2023accurate, lam2023graphcast} and probabilistic models~\cite{oskarsson2024probabilistic, price2025probabilistic, alet2025skillful, lang2026aifs}, could offer new ways to achieve high-fidelity, data-driven plasma surrogates. In particular, we adapt GNN architectures~\cite{keisler2022forecasting, oskarsson2024probabilistic} to emulate plasma dynamics in the noon–midnight meridional plane of the Earth's magnetosphere. We specifically focus on the sensitivity of the magnetospheric response to upstream conditions by compiling a dataset of four global 2D + 3V Vlasiator simulations. In these runs, the solar wind ion density is systematically increased, effectively scanning a range of Alfvén Mach numbers from moderately to strongly super-Alfvénic regimes. Our GNN emulators are trained on the electromagnetic fields and plasma moments of this density-varied dataset, achieving a speedup of more than two orders of magnitude on a single GPU compared to the original simulation on 100 CPUs. The framework supports both deterministic forecasting and probabilistic ensemble generation through a latent-variable formulation that yields forecast uncertainty. Rather than targeting a specific operational space-weather index or research diagnostic, we treat the problem as a faithful-reproduction benchmark against the simulator. The aim is to establish uncertainty-aware graph-based emulation of hybrid-Vlasov moment fields, diagnose where it succeeds and fails, and release the dataset~\cite{vlasiator2025mldata} and code~\cite{spacecast2025} as an open benchmark for further machine learning studies of ion-kinetic magnetospheric dynamics. The rest of the paper is organized as follows. Section~2 describes the hybrid-Vlasov dataset. Section~3 introduces the Graph-FM and Graph-EFM models. Section~4 details the experimental setup. Section~5 reports the forecast results. Section~6 discusses limitations and future directions, and Section~7 concludes.

\section{Hybrid-Vlasov dataset}

The data for this study were generated using Vlasiator, which performs global simulations of the solar wind's interaction with the Earth's magnetosphere in the hybrid-Vlasov formalism. Ions are treated as a velocity distribution function (VDF), $f$, that depends on position $\mathbf{x}$, the velocity-space coordinate $\mathbf{v}$, and continuous time $t$. Its evolution is dictated by the Vlasov equation, which describes how $f$ changes due to the electric field $\mathbf{E}$ and magnetic field $\mathbf{B}$:
\begin{equation}
\frac{\partial f}{ \partial t} + {\bf v} \cdot \frac{\partial f}{\partial {\bf x} } + \frac{q_i}{m_i} \left( {\bf E} + {\bf v} \times {\bf B} \right) \cdot \frac{\partial f}{\partial {\bf v } } = 0,
\end{equation}
where $q_i$ and $m_i$ denote the ion charge and mass. We use $\mathbf{v}$ for the velocity-space coordinate and $\mathbf{V}$ for the ion bulk velocity, the first moment of $f$, which appears below. Electrons are modeled as a massless, charge-neutralizing fluid where the number density $n$ is equal for both ions and electrons ($n_i \simeq n_e \simeq n$). The electromagnetic fields are evaluated by solving the Maxwell-Darwin system:
\begin{equation}
\nabla \times \mathbf{E} = -\partial_t \mathbf{B},\ \nabla \times \mathbf{B} = \mu_0 \mathbf{J},\ \nabla \cdot \mathbf{B} = 0.
\end{equation}
This system is closed by relating the fields to the moments of the ion distribution function through a generalized Ohm's law that includes the Hall term, which involves the ion bulk velocity $\mathbf{V}$, the current density $\mathbf{J}$ and the elementary charge $e$:
\begin{equation}
{\bf E} + {\bf V} \times {\bf B} = \frac{ {\bf J} \times {\bf B} }{n e}.
\end{equation}

All simulations are performed in 2D + 3V (two spatial and three velocity dimensions) in the noon–midnight meridional plane in the Geocentric Solar Ecliptic (GSE) coordinate system. The spatial grid is static, Cartesian and uniformly spaced with the resolution of $600\,\mathrm{km}$. The spatial domain spans $x\in[-60,30]\,R_E$ and $z\in[-30,30]\,R_E$. The velocity space is also represented by a Cartesian grid, with a uniform resolution of $52\,\mathrm{km}/\mathrm{s}$. The temporal cadence of the simulations is $dt = 0.035\,\mathrm{s}$. The cadence of the reduced output files is $\Delta t = 1\,\mathrm{s}$. The inner boundary at $r=3.7\,R_E$ is modeled as a perfect conductor, while the dayside inflow boundary injects a Maxwellian solar wind. The outflow boundaries at $\pm z$ and $-x$ employ copy conditions, while the $y$ direction is treated as a periodic dimension with a single layer of cells. 

The simulations are carried out in the ($x$–$z$) plane at $y = 0$. This plane lies along the Sun–Earth line and includes the geomagnetic dipole axis, and Vlasiator simulations in this configuration have been shown to capture key features such as the collisionless bow shock, ion foreshock, and magnetosheath \cite{vonalfthan2014vlasiator}, as well as magnetic reconnection at the dayside magnetopause \cite{hoilijoki2017reconnection} and in the magnetotail~\cite{palmroth2017tail}. Restricting the simulations to two spatial dimensions implies an out-of-plane symmetry with no gradients in that direction, i.e.\ $\partial/\partial y = 0$. As a result, large-scale transport of magnetic flux around the magnetosphere via the dawn--dusk flanks is suppressed, and in configurations with southward interplanetary magnetic field the solar-wind flux must reconnect directly on the dayside. While this removes certain global transport pathways, the 2D + 3V approximation remains physically meaningful for many magnetospheric processes.

To probe the parameter space, we vary the solar wind ion density while keeping the velocity, temperature, and interplanetary magnetic field fixed. As shown in Table~\ref{tab:runs}, increasing the density systematically raises the Alfvén Mach number $M_A$, moving the system from moderately super-Alfvénic (Run~1, $M_A=4.9$) to strongly super-Alfvénic flow (Run~4, $M_A=9.8$). Because the bow shock standoff distance and the overall magnetospheric morphology depend sensitively on $M_A$, this controlled sweep provides a dataset spanning a range of relevant physical regimes. The physical cell size is held fixed at $600\,\mathrm{km}$, so the same sweep also changes the resolved kinetic scale. In hybrid-kinetic convergence studies the grid spacing is commonly kept fixed relative to the ion inertial length $d_i$ when the density is changed, so that the kinetic physics is resolved equally well across runs. Here the upstream ion inertial length instead decreases from ${\sim}320$\,km in Run~1 to ${\sim}160$\,km in Run~4, and $\Delta x / d_i$ increases from roughly 1.9 to 3.7. The four runs therefore do not form a numerical-convergence sequence. Differences between them reflect both changed upstream parameters and changed kinetic-scale resolution relative to $d_i$~\cite{dubart2020resolution}. The emulator is trained to reproduce the Vlasiator solutions as they are, and the sweep is a controlled emulation benchmark across qualitatively distinct magnetospheric states rather than a physical parameter scan at fixed numerical fidelity.

\begin{table}[ht]
\centering
\caption{Solar wind parameters at the dayside boundary, and time step information for each simulation run.}
\label{tab:runs}
\begin{tabular}{lccccccc}
\toprule
Label & $\rho$ (cm$^{-3}$) & $\boldsymbol{V}$ (km/s) & $\boldsymbol{B}$ (nT) & $T$ (MK) & $M_A$ & $\Delta t$ (s) & $t_{\mathrm{tot}}$ (s) \\
\midrule
Run 1 & 0.5  & $(-750,\ 0,\ 0)$ & $(0, 0, -5)$ & $0.5$ & 4.9 & 1.0 & 800 \\
Run 2 & 1.0  & $(-750,\ 0,\ 0)$ & $(0, 0, -5)$ & $0.5$ & 6.9 & 1.0 & 800 \\
Run 3 & 1.5  & $(-750,\ 0,\ 0)$ & $(0, 0, -5)$ & $0.5$ & 8.4 & 1.0 & 800 \\
Run 4 & 2.0  & $(-750,\ 0,\ 0)$ & $(0, 0, -5)$ & $0.5$ & 9.8 & 1.0 & 800 \\
\bottomrule
\end{tabular}
\end{table}

We use a set of plasma and electromagnetic variables derived from Vlasiator simulations, including magnetic ($B_x$, $B_y$, $B_z$) and electric ($E_x$, $E_y$, $E_z$) field components, bulk velocity components ($V_x$, $V_y$, $V_z$), particle number density ($\rho$), plasma pressure ($P$), and temperature ($T$). These correspond to standard moments of the ion velocity distribution function. The residual standard deviations are used to weight the loss function, normalizing for differences in dynamical variability across variables. The data is stored in Zarr format~\cite{miles2024zarr} in an open repository \cite{vlasiator2025mldata}. A detailed summary of all variables, including their notation, units, and residual standard deviations across the four simulation runs, is provided in Appendix~\ref{app:data}.

\section{Methods}

\subsection{Problem formulation}
\begin{figure}[ht]
  \centering
  \includegraphics[width=\textwidth]{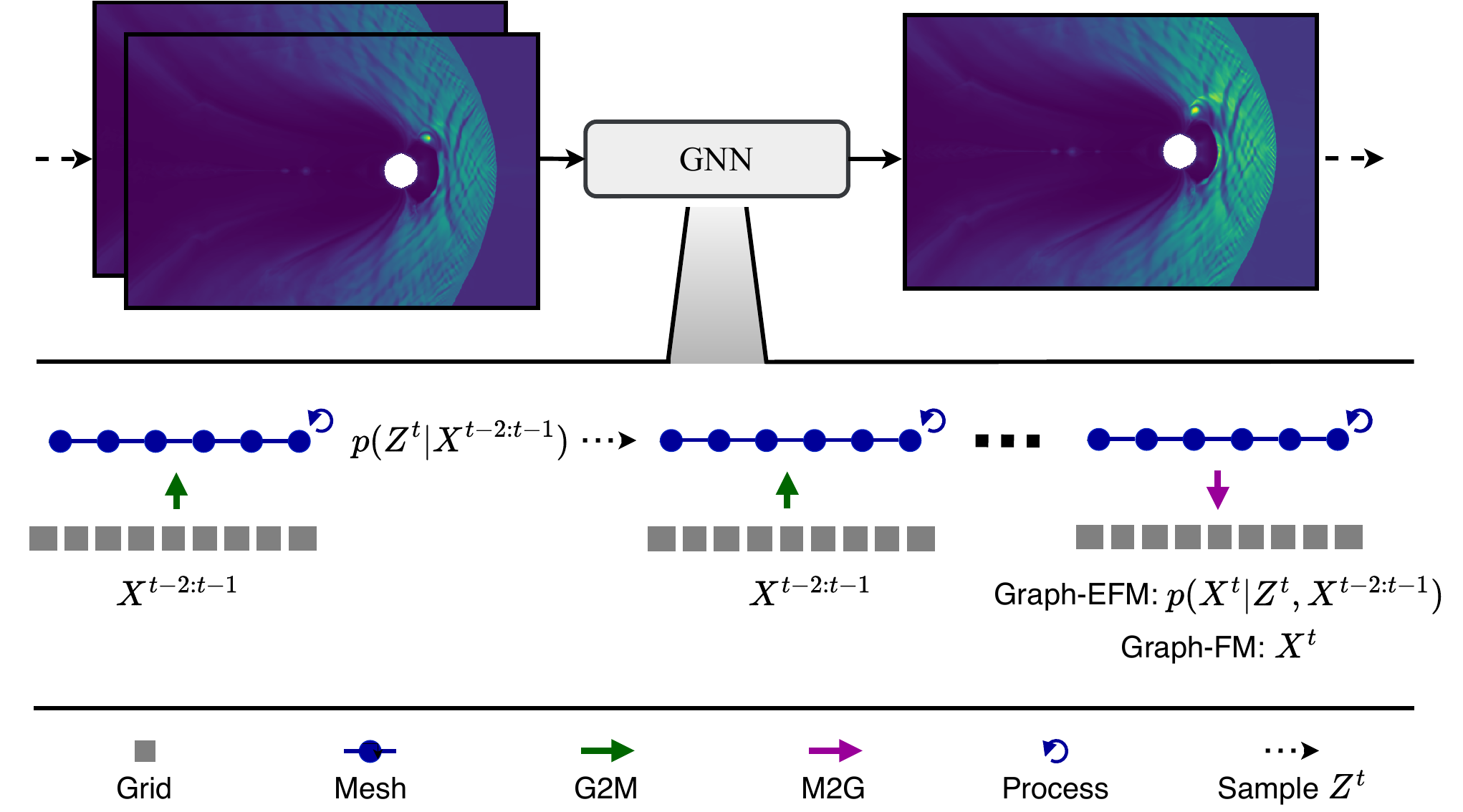}
  \caption{Schematic overview of the Graph-FM/EFM forecasting framework. Top: the two most recent simulation states (particle number density depicted here) are mapped to the next state by a graph neural network (GNN); dashed arrows indicate that predicted states are fed back, together with the most recent input, so the mapping can be applied autoregressively to produce longer rollouts. Middle: expanded view of that mapping. Grid states $X^{t-2:t-1}$ are encoded onto a coarse mesh, processed by message passing, and decoded back to the grid. Graph-FM is the deterministic encode--process--decode path that outputs $X^t$. Graph-EFM uses the same predictor with an additional latent map $p(Z^t \mid X^{t-2:t-1})$; a sample of $Z^t$ is injected into the mesh so that $p(X^t \mid Z^t, X^{t-2:t-1})$ generates one ensemble member. Bottom: legend for grid and mesh nodes, grid-to-meh (G2M) and mesh-to-grid (M2G) edges, the process step, and sampling of $Z^t$.}
  \label{fig:schema}
\end{figure}

We formulate the problem of space plasma forecasting as mapping a set of initial magnetospheric states to a sequence of future states as shown in Figure~\ref{fig:schema}. The input consists of two consecutive states, $X^{-1:0} = (X^{-1}, X^{0})$, to capture first-order dynamics~\cite{lam2023graphcast}, with the goal of predicting the future trajectory $X^{1:H} = (X^{1}, \dots, X^{H})$. A superscript $t$ is a discrete time-step index at the 1\,s output cadence, distinct from the continuous physical time in the Vlasov equation, and $H$ is the forecast horizon. Each magnetospheric state $X^t \in \mathbb{R}^{N \times F}$ is a high-dimensional tensor representing $F$ physical variables present in the hybrid-Vlasov simulation across $N$ grid locations in near-Earth space. Deterministic models tackle the problem by producing a single point, typically mean, estimate of the trajectory $X^{1:H}$, while probabilistic approaches aim to model the full conditional distribution $p(X^{1:H} | X^{-1:0})$.

\subsection{Graph-based neural forecasting}
Using machine learning, the deterministic forecasting problem can be solved with an \emph{autoregressive} model, where predictions are made iteratively by using the model's previous output as input for the next time step. In the probabilistic case, we sample from the model's output distribution and repeat the process to generate an ensemble of trajectories. For both cases, we use a GNN based on an \emph{encode-process-decode} architecture~\cite{sanchez2020learning}. In the encoding stage, the physical variables defined on the high-resolution simulation grid are projected onto a coarser mesh, where each mesh node receives information from its corresponding grid neighborhood. During the processing stage, several GNN message-passing layers operate on this mesh graph, allowing information to propagate laterally across mesh nodes and expanding each node’s receptive field. In the decoding stage, the updated mesh features are mapped back to the original grid via directed mesh-to-grid edges to produce the next predicted state. The GNN outputs a \emph{residual update} to the most recent input state, making it an easier learning task compared to predicting the next state directly.

We construct our mesh graph $\mathcal{G}_\mathrm{M}$ for our forecasting models by downsampling the simulation grid. This produces a quadrilateral mesh structure~\cite{oskarsson2024probabilistic} coarser than the simulation domain~\cite{keisler2022forecasting} that is computationally feasible to learn next step prediction on. Information flows between representations via three edge sets: grid-to-mesh, with edges directed from grid nodes to mesh nodes; mesh-to-mesh, whose edges are bidirectional and support lateral message passing within the mesh; and mesh-to-grid, with edges directed from mesh nodes back to grid nodes.

To ensure accurate handling of open boundary conditions at the solar wind inflow region on the Earth's dayside one could apply a boundary forcing at the domain edge by specifying which grid nodes are to be replaced with ground truth boundary data after each prediction~\cite{holmberg2025graph}. However, since the simulations in this work are driven by four distinct but constant solar wind conditions, this is not strictly necessary, as the model is given all the information necessary to learn the corresponding steady inflow for a given initial state without explicit boundary forcing. For experiments involving time-varying boundary conditions, this effect should be modeled explicitly.

\subsection{Deterministic model}

We use the deterministic graph-based forecasting model \emph{Graph-FM}~\cite{oskarsson2024probabilistic} to generate point estimate forecasts through the autoregressive mapping $\hat{X}^{t} = f(X^{t-2:t-1})$. GNN message passing is performed to encode information from the grid to the coarser mesh, then a number of GNN message passing processing steps takes place on this coarser mesh, and a final decoding using GNNs is performed to map back to the next simulation grid state. All upward updates are facilitated by \emph{propagation networks} \cite{oskarsson2024probabilistic}, while the remaining updates use \emph{interaction networks}~\cite{battaglia2016interactionnet}, with all layers mapping to a latent dimensionality $d_z$. This design leverages the inductive bias of interaction networks to retain information, while propagation networks are more effective at forwarding new information through the graph. We train the deterministic models by minimizing a weighted mean square error (MSE) loss.

\subsection{Probabilistic model} For probabilistic forecasting we employ the graph-based ensemble forecasting model, \emph{Graph-EFM}~\cite{oskarsson2024probabilistic}. To model the distribution $p(X^{t}|\arcond)$ for a single time step, Graph-EFM uses a latent random variable $Z^t$. This variable acts as a low-dimensional representation of the stochastic elements of the system that are not captured by the input states. By conditioning the prediction on this latent variable, the model can efficiently sample different, spatially coherent outcomes. The relationship is defined by the integral:
\begin{equation}
p(X^{t}|\arcond) = \int p(X^{t}|Z^{t}, \arcond) p(Z^{t}|\arcond) dZ^{t}
\end{equation}
where the term $p(Z^{t}|\arcond)$ is the latent map and the term $p(X^{t}|Z^{t}, \arcond)$ is the predictor. The latent map is a probabilistic mapping that takes the previous two states as input and produces the parameters for a distribution over the latent variable $Z^t$. Specifically, Graph-EFM uses GNNs to map the inputs to the mean of an isotropic Gaussian distribution, effectively encoding the state-dependent uncertainty into the latent space. The variance is kept fixed. This latent distribution is defined over the nodes $\mathcal{V}_\textit{M}$ in the coarse mesh graph, as follows:
\begin{equation}
p(Z^{t}|\arcond) = \prod_{v \in \mathcal{V}_\textit{M}} \mathcal{N}(Z_{v}^{t} | \mu_{Z}(\arcond)_{v}, I),
\end{equation}
 ensuring that the stochasticity is introduced in a lower-dimensional, spatially-aware way. The predictor is a deterministic mapping that takes a specific sample of the latent variable $Z^t$, along with the previous states, to produce the next state $\hat{X}^t$. The predictor is a deterministic mapping, $g$, that produces the next state $\hat{X}^t$ by adding a predicted residual update, $\tilde{g}$, to the previous state $X^{t-1}$:
\begin{equation}
\hat{X}^{t} = g(Z^{t}, \arcond) = X^{t-1} + \tilde{g}(Z^{t}, \arcond).
\end{equation}
By first sampling a $Z^t$ from the latent map and then passing it through the predictor, the model generates one possible future state. Repeating this process creates an arbitrarily large ensemble of forecasts. Drawing from the structure of a conditional Variational Autoencoder (VAE)~\cite{kingma2014vae, sohn2015cvae}, we train the Graph-EFM by optimizing a variational objective equivalent to the negative Evidence Lower Bound (ELBO), which combines a Kullback-Leibler (KL) divergence regularizer with a reconstruction loss. Subsequently, we fine-tune the model by adding a Continuous Ranked Probability Score (CRPS) loss~\cite{gneiting2007strictly, rasp2018neural, pacchiardi2024probabilistic} to the objective, for calibration.

\subsection{Model details}

The $1006 \times 671\,(x, z)$ data grid excludes $5124$ inner boundary nodes close to the Earth. The mesh used by the models is constructed by placing each mesh node at the center of a $5\times 5$ square of finer simulation grid cells. Because the mesh does not cover the inner boundary, the GNN naturally excludes this region from the computation, unlike e.g. convolutional networks which would require masking in this instance. The full statistics of this graph including graph-to-mesh and mesh-to-graph connections are shown in Table~\ref{tab:graphs}. For each node, a multilayer perceptron (MLP) encodes static features in Table~\ref{tab:dataset}, concatenated with previous states. For a complete explanation on update rules in the GNNs see~\cite{oskarsson2024probabilistic}. All MLPs use one hidden layer with Swish activation and layer normalization.

\begin{table}[h]
\centering
\caption{Number of nodes and edges in each graph, where $\mathcal{G}_\mathrm{M}$ is the mesh-to-mesh graph, $\mathcal{G}_\mathrm{G2M}$ the grid-to-mesh graph, and $\mathcal{G}_\mathrm{M2G}$ the mesh-to-grid graph.}
\begin{tabular}{lrr}
\toprule
Graph & Nodes & Edges \\
\midrule
$\mathcal{G}_\mathrm{M}$ & 15496 & 122344 \\
$\mathcal{G}_\mathrm{G2M}$    & -       & 669902 \\
$\mathcal{G}_\mathrm{M2G}$    & -       & 2009706 \\
Grid & 669902   & - \\
\bottomrule
\end{tabular}
\label{tab:graphs}
\end{table}

Some modifications to the model were necessary to enable training on our substantially larger grid of $669{,}902$ nodes, compared to $29{,}040$ grid nodes for global weather and $63{,}784$ for limited-area weather using hierarchical GNNs~\cite{oskarsson2024probabilistic}. First, we construct the meshes using a coarser $5\times5$ downsampling ratio instead of $3\times3$, reducing the number of nodes in the mesh graph. Second, in the mesh-to-grid mapping stage, each grid node is connected to its three nearest mesh nodes rather than four to reduce the size of the mapping with the largest number of edges in the architecture, constituting a main memory bottleneck, while still providing an interpolation stencil spanning the plane. Third, we introduce an additional MLP prior to the mesh-to-grid projection, defining an intermediate dimension $d_{\mathrm{M2G}}$ as this directly affects the largest matrix multiplication in the model. These design choices allow us to increase the latent dimension while still being able to fit the model in GPU memory, and empirically they also do not seem to produce visual artifacts. Finally, we apply gradient checkpointing~\cite{chen2016training} across rollout steps, ensuring that GPU memory usage remains effectively constant when increasing rollout length during training, while compute time scales approximately linearly with the number of steps. The code is implemented in PyTorch Lightning (v2.4.0)~\cite{falcon2024lightning} and made openly available~\cite{spacecast2025}.

\subsection{Deterministic objective}

We train the deterministic models by minimizing a weighted MSE:
\begin{equation}
\mathcal{L}_\text{MSE} = \frac{1}{HN} \sum_{t=1}^H \sum_{n=1}^N \sum_{i=1}^{d_x} \frac{\lambda_i}{d_x} \left( \hat{X}^t_{n, i} - X^t_{n, i} \right)^2.
\label{eq:mse}
\end{equation}
The loss weight $\lambda_i$ is the inverse variance of the time differences for variable $i$, which normalizes the contribution of variables with different dynamic ranges~\cite{keisler2022forecasting}.

To encourage physical consistency, we augment the deterministic training loss with a divergence penalty enforcing $\nabla \cdot \mathbf{B} = 0$. The divergence loss is defined as
\begin{equation}
\label{eq:div_loss}
\divloss
= \frac{1}{HN} \sum_{t=1}^H \sum_{n=1}^{N}
\left(
\frac{\partial \hat{B}^t_x}{\partial x}
+
\frac{\partial \hat{B}^t_z}{\partial z}
\right)_n^2,
\end{equation}
where the divergence $\nabla \cdot \hat{\mathbf{B}} = \partial_x \hat{B}_x + \partial_z \hat{B}_z$ is computed over the interior grid points. Analogous to how divergence may accumulate over time in MHD simulations~\cite{janhunen2012gumics} and is removed using projection methods~\cite{brackbill1980effect}, we optimize the soft constraint over multiple rollout steps $H$ to prevent long-term buildup of divergence. The spatial derivatives are discretized using second–order central differences,
\begin{align}
\left.\frac{\partial \hat{B}_x}{\partial x}\right|_{i,j}
&\approx
\frac{\hat{B}_x(x_{i+1}, z_j) - \hat{B}_x(x_{i-1}, z_j)}{2\,\Delta x}, \\
\left.\frac{\partial \hat{B}_z}{\partial z}\right|_{i,j}
&\approx
\frac{\hat{B}_z(x_i, z_{j+1}) - \hat{B}_z(x_i, z_{j-1})}{2\,\Delta z},
\end{align}
yielding second–order accurate estimates on the interior stencil.  
The final deterministic training objective becomes
\begin{equation}
\mathcal{L}
= \mathcal{L}_\mathrm{MSE} + \lambda_\mathrm{Div} \,\divloss,
\end{equation}
where $\lambda_{\mathrm{Div}}$ controls the strength of the divergence penalty.

\subsection{Probabilistic objective}

The probabilistic model's single-step prediction has a structure analogous to a conditional VAE, and it is trained by optimizing a variational objective derived from the ELBO:
\begin{align}
    \begin{split}
    \label{eq:elbo_loss}
    \tilde{\mathcal{L}}_{\mathrm{Var}}&\left(X^{t-2:t-1}, X^t \right)
    =
    \lambda_{\mathrm{KL}} D_{\mathrm{KL}}\left({\vardist }\middle|\middle|\,{\priordist}\right)
    \\
    -&\mathbb{E}_{\vardist}\left[{
        {\textstyle \sum_{n=1}^N
        \sum_{i = 1}^{d_x}}
        \log \mathcal{N}\left({
            X^{t}_{v, i}
        }\middle|{
            g\left(Z^t, \arcond\right)_{v, i}
        },{
            \sigma^2_{v, i}
        }\right)
    }\right].
    \end{split}
\end{align}
Here, the variational distribution $q\left(Z^t \middle| X^{t-2:t-1}, X^t\right)$ provides a learned approximation to the intractable true posterior $p\left(Z^t \middle| X^{t-2:t-1}, X^t\right)$ over the latent variables $Z^t$. This distribution is parameterized in a similar way as the latent map used for forecasting: a set of GNN layers encodes the inputs and produces the mean and variance of a Gaussian distribution over $Z^t$. Unlike the prior $p\left(Z^t \middle| X^{t-2:t-1}\right)$, the variational approximation $q$ is posterior-facing, i.e. it also conditions on the target state $X^t$ in order to match the true posterior as closely as possible, consistent with standard variational inference. This objective consists of two terms. The first is the KL divergence, which acts as a regularizer, encouraging the approximate posterior $q$ to remain close to the prior $p$. The second term is the expected negative log-likelihood, or reconstruction loss, which trains the predictor $g$ to accurately reconstruct the true state $X^t$ from a latent sample $Z^t$. The hyperparameter $\lambda_\mathrm{KL}$ balances these terms to prevent the model from collapsing to a deterministic prediction. Because both distributions are Gaussian, the KL divergence has a closed-form solution. The predictive variance $\sigma^2_{v,i}$ is produced by the decoder network.

As in the deterministic setting, we roll out the forecasts over multiple time steps to promote long-term stability. The multi-step objective is defined as:
\begin{equation}
\varloss = \sum_{t=1}^{H} \tilde{\mathcal{L}}_{\mathrm{Var}} \left(\hat{X}_{t-2:t-1},\, X_t\right) \label{eq:variational_finetune}
\end{equation}
where $\hat{X}_t$ denotes an initial state from the dataset for $t<1$, and for $t\ge 1$ is obtained by sampling from the variational approximation of $Z^t$.

The model is further fine-tuned using the CRPS loss, which is minimized only when the predicted distribution matches the empirical data distribution. We use an unbiased two-sample estimator for the CRPS loss, summed over all grid points and variables:
\begin{equation}
    \label{eq:crps_loss}
    \crpsloss = \frac{1}{HN}\sum_{t=1}^H\sum_{n=1}^N \sum_{i = 1}^{d_x}
    \frac{1}{2} \left(
        \left|\hat{X}^t_{n , i} - X^t_{n, i}\right|
        + \left|\check{X}^t_{n, i} - X^t_{n, i}\right|
        - \left|\hat{X}^t_{n, i} - \check{X}^t_{n, i}\right|
    \right)
\end{equation}
where $\hat{X}^t$ and $\check{X}^t$ are two independent ensemble members (forecasts) generated by the model. Similarly to the deterministic case, we add a divergence loss term as in Eq.~\eqref{eq:div_loss} computed on the predicted $\hat{B}_x$ and $\hat{B}_z$ components. The final training loss is the combination of these different objectives:
\begin{equation}
\mathcal{L} = \varloss + \lambda_\mathrm{CRPS} \crpsloss + \lambda_\mathrm{Div} \divloss.
\end{equation}

Computing the full probabilistic training loss requires rolling out three separate forecasts: one for the variational objective $\varloss$ (using latent samples drawn from the variational distribution $q$) and two for the CRPS objective $\crpsloss$ (using samples from the latent map). This introduces a substantial computational overhead, but in practice the cost is manageable because the CRPS term is only included during the final stages of training.

\section{Experiments}

To evaluate our models, the Vlasiator simulations are causally split into training, validation, and test sets with durations of 12 minutes, 20 seconds, and 1 minute, respectively. This chronological split has no temporal overlap between training and test periods, so the evaluation tests temporal generalization, i.e. how well the models can predict the future of each run. All four runs appear in all three splits, and the test is therefore within the upstream regimes seen during training. During rollout training, sequences are sampled using a windowed approach: a fixed-length temporal window of consecutive timesteps is selected from within a split, and the model predicts the next step(s) within that window. Each window is fully contained within its split, so no rollout sequence crosses the boundary between training, validation, or test sets. We train both the deterministic Graph-FM and the probabilistic Graph-EFM models. All models are configured with latent dimensions $d_z = 256$ and $d_{\mathrm{M2G}}=128$. The processor consists of 12 processing layers. Graph-EFM is set to generate an ensemble size of 5. The models produce 30\,s long forecasts during evaluation initialized from each simulation step in the test loader so that there are sufficient samples to calculate evaluation statistics on.

\subsection{Training details}
\label{sec:training}

All models are trained using the AdamW optimizer~\cite{loshchilov2019adamw} with parameters $\beta_1 = 0.9$, $\beta_2 = 0.95$, a weight decay of $0.01$, and an effective batch size of 32. The models are trained using bfloat16 mixed precision to save GPU memory. The full training schedules for both deterministic and probabilistic models are summarized in Table~\ref{tab:schedule}. The deterministic Graph-FM model is trained in three phases as detailed in the upper section of Table~\ref{tab:schedule}. The first 175 epochs use a learning rate of $10^{-3}$ with single-step unrolling. This is followed by 50 epochs at a reduced learning rate of $10^{-4}$ while gradually increasing the rollout length from 1 to 4 steps. We set the maximum rollout length to 4 steps based on prior work in weather emulation~\cite{keisler2022forecasting}, which shows that increasing the number of unrolled steps improves long-term predictive performance, but with diminishing returns beyond 4 steps. It is more expensive to train with a larger number of rollout steps because the forward pass and loss have to be computed for each additional step, but some works like GraphCast~\cite{lam2023graphcast} increase the number of steps up to 12 when training. We selected 4 steps as a computationally feasible value, and most of the training uses single-step prediction, which is much faster. The cap reflects computational cost rather than an observed stability threshold. A final 25-epoch stage continues with $H = 4$ and introduces the magnetic divergence penalty with weight $\lambda_\mathrm{Div} = 10$. The value for $\lambda_\mathrm{Div}$ is chosen such that $\divloss$ decreases while the validation error of the predicted $B_x$ and $B_z$ components do not increase. This loss term is added only at the final stage so we can compare performance before and after its inclusion and explore different weight values at lower computational cost.

\begin{table}[ht]
\centering
\caption{Training schedules for the deterministic and probabilistic models.}
\label{tab:schedule}
\begin{tabular}{lcccccc}
\toprule
Model & Epochs & Learning Rate & Unrolling $H$ & $\lambda_\text{KL}$ & $\lambda_\text{CRPS}$ & $\lambda_\text{Div}$ \\
\midrule
\multirow{3}{*}{Graph-FM} & 175 & $10^{-3}$ & 1 & - & - & 0 \\
 & 50 & $10^{-4}$ & 1--4 & - & - & 0 \\
 & 25 & $10^{-4}$ & 4 & - & - & 10 \\
\midrule
\multirow{5}{*}{Graph-EFM} & 100 & $10^{-3}$ & 1 & 0 & 0 & 0 \\
 & 50 & $10^{-3}$ & 1 & 1 & 0 & 0 \\
 & 50 & $10^{-4}$ & 1--4 & 1 & 0 & 0 \\
 & 25 & $10^{-4}$ & 4 & 1 & $10^6$ & 0 \\
 & 25 & $10^{-4}$ & 4 & 1 & $10^6$ & $10^7$ \\
\bottomrule
\end{tabular}
\end{table}

The probabilistic Graph-EFM model is trained in five stages as shown in the lower section of Table~\ref{tab:schedule}. In the first stage, we pretrain the network as a deterministic autoencoder by setting $\lambda_{\mathrm{KL}} = 0$ and $\lambda_\mathrm{CRPS} = 0$ for 100 epochs. This encourages $q$ to encode meaningful information in the distribution over $Z^t$, and helps prevent the model from ignoring the latent variable and collapsing to deterministic forecasting~\cite{oskarsson2024probabilistic}. In the second stage, we activate the variational term with $\lambda_{\mathrm{KL}} = 1$ and train for 50 epochs on single-step predictions. The number of epochs is selected to give the training enough time to reach convergence. The third stage reduces the learning rate from $10^{-3}$ to $10^{-4}$ and increases the unroll length from 1 to 4 steps over 50 epochs to promote temporal stability. In the fourth stage, we introduce the CRPS loss with $\lambda_\mathrm{CRPS} = 10^{6}$ for an additional 25 epochs. Finally, we add the magnetic divergence penalty with $\lambda_\mathrm{Div} = 10^{7}$ for a 25-epoch fine-tuning stage. The $\lambda_\mathrm{CRPS}$ is chosen so that the Spread--Skill Ratio (SSR) in Eq.~\eqref{eq:ssr} clearly increases without getting any visual artifacts, and the divergence weight is chosen in a similar way to the deterministic model so as to not increase the validation error for $B_x$ and $B_z$ while minimizing $\nabla \cdot \mathbf{B}$.

\subsection{Computational complexity}

Table~\ref{tab:model_times} summarizes the training and inference performance of the proposed models. Each training phase was conducted on 32~AMD~MI250X GPUs. The CRPS stage (the final 50~epochs of Graph-EFM training) is by far the most computationally demanding training phase, requiring over 70\,h for each model configuration. This is because computing the loss then requires rolling out two additional forecasts. Inference was benchmarked on a single~AMD~MI250X GPU and measured as the time it takes to complete one next-step prediction. For comparison, the physics-based Vlasiator simulation requires approximately 4--5\,minutes on 100~AMD~EPYC~7H12~CPUs with 64 cores each to simulate 1\,s of physical time.

On a single GPU, the deterministic models predict the next step approximately 160 times faster than the original simulation running on 100 CPUs, while our probabilistic models are around 20 times faster with an ensemble size of 5. These figures are indicative wall-time ratios, not a like-for-like benchmark. The comparison uses different hardware, one GPU versus 100 CPUs, and different computed quantities. The emulators advance only the moments of the ion VDF at a 1\,s cadence, whereas Vlasiator advances the full VDF with a much smaller internal time step. The numbers therefore describe the practical cost of generating moment-field forecasts, not an equal-fidelity acceleration of the kinetic solver.

\begin{table}[ht]
\centering
\caption{Training and inference wall times for the Graph-FM and Graph-EFM models. Training was performed on 32~AMD~MI250X GPUs, while inference times correspond to a single~GPU. For Graph-EFM, the training time in parentheses indicates the time before CRPS fine-tuning, and inference times are reported per ensemble member.}
\label{tab:model_times}
\begin{tabular}{lccc}
\toprule
Model & Parameters (M) & Training time (h) & Inference time (s) \\
\midrule
\multirow{1}{*}{Graph-FM} & 6.6 & 39.6 & 1.67 \\
\multirow{1}{*}{Graph-EFM} & 9.0 & 109 (35.3) & 2.46 \\
\bottomrule
\end{tabular}
\end{table}

\subsection{Metrics}
\label{sec:metrics}

To evaluate the performance of our forecasts, we use a set of standard metrics. For an ensemble forecast with $M$ members (deterministic forecasts have $M=1$), we denote the prediction for variable $i$ at location $n$ for a given sample $s$ at time $t$ as $\hat{X}_{n,i}^{s,t,m}$, with the corresponding ground truth being $X_{n,i}^{s,t}$. The Root Mean Squared Error (RMSE) is calculated by averaging the squared error over all $S$ forecasts in the test set and all $N$ grid locations:
\begin{align}
\mathrm{RMSE}_{t,i} &= \sqrt{\frac{1}{SN}\sum_{s=1}^{S}\sum_{n=1}^N \left(\overline{X}_{n,i}^{s,t} - X_{n,i}^{s,t}\right)^{2}} \\
\text{where} \quad \overline{X}_{n,i}^{s,t} &= \frac{1}{M}\sum_{m=1}^{M}\hat{X}_{n,i}^{s,t,m}.
\end{align}

The CRPS is calculated for ensemble forecasts to assess the overall quality of a probabilistic forecast by comparing the entire predictive distribution to the single ground truth observation. Lower values indicate better performance. We use a finite-sample estimate~\cite{zamo2018estimation}, computed as:
\begin{align}
\label{eq:crps}
\begin{split}
\mathrm{CRPS}_{t,i} = \frac{1}{SN}\sum_{s=1}^{S}\sum_{n=1}^N \Biggl( &\frac{1}{M}\sum_{m=1}^{M}\left|\hat{X}_{n,i}^{s,t,m} - X_{n,i}^{s,t}\right| \\
&-\frac{1}{2M(M-1)}\sum_{m=1}^{M}\sum_{m'=1}^{M}\left|\hat{X}_{n,i}^{s,t,m} - \hat{X}_{n,i}^{s,t,m'}\right| \Biggr).
\end{split}
\end{align}

The ensemble spread quantifies the forecast uncertainty represented by the ensemble. It is defined as the root-mean-square deviation of the ensemble members from their ensemble mean:
\begin{equation}
\mathrm{Spread}_{t,i} = \sqrt{\frac{1}{SMN}\sum_{s=1}^{S}\sum_{m=1}^{M}\sum_{n=1}^N \left(\overline{X}_{n,i}^{s,t} - \hat{X}_{n,i}^{s,t,m}\right)^{2}}
\end{equation}
From this, the bias-corrected SSR is defined as
\begin{equation}
\label{eq:ssr}
\mathrm{SSR}_{t,i} = \sqrt{\frac{M+1}{M}} \frac{\mathrm{Spread}_{t,i}}{\mathrm{RMSE}_{t,i}}
\end{equation}
An ensemble is considered well-calibrated when $\mathrm{SSR} \approx 1$, indicating that the ensemble spread accurately reflects the forecast error~\cite{fortin2014should}.

\section{Results}

\subsection{Example forecasts}

\begin{figure}[ht]
\centering
\includegraphics[width=\textwidth]{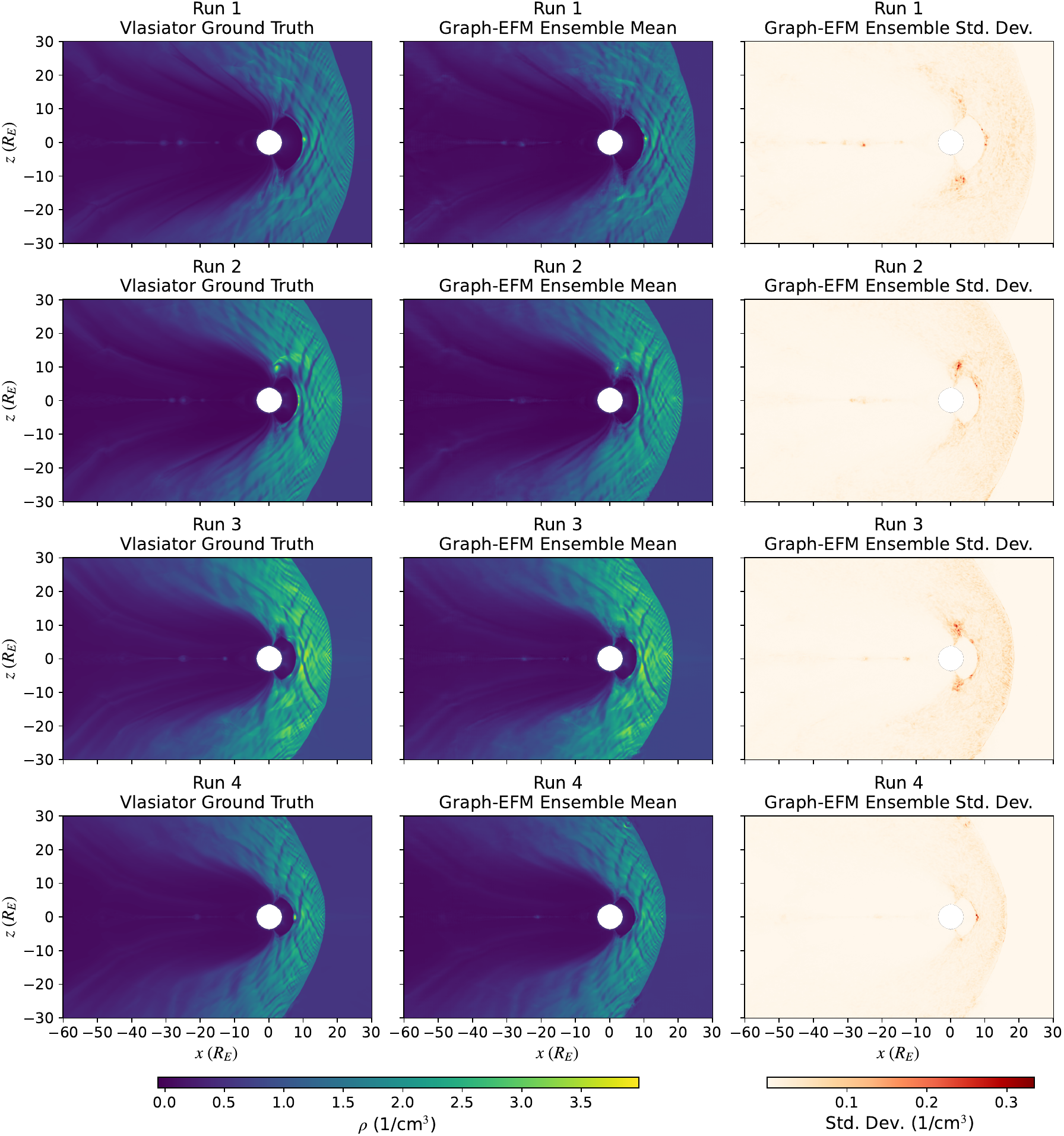}
\caption{Example forecasts of plasma density $\rho$ produced by Graph-EFM at lead time $t=30\,\mathrm{s}$ for the four Vlasiator simulation runs. Rows correspond to Runs 1--4 with increasing upstream solar wind ion density ($0.5$, $1.0$, $1.5$, and $2.0~\mathrm{cm^{-3}}$). Columns show the Vlasiator ground truth, the Graph-EFM ensemble mean forecast, and the ensemble standard deviation.}
\label{fig:rho}
\end{figure}

Figure~\ref{fig:rho} presents example Graph-EFM forecasts for the plasma density $\rho$ from four distinct Vlasiator runs, corresponding to solar wind ion densities of $0.5$, $1.0$, $1.5$, and $2.0~\mathrm{cm^{-3}}$ (rows 1–4). As the upstream density increases, the bow shock is progressively compressed toward Earth. The model is trained to learn all four runs, and manages to successfully roll out physically consistent forecasts for each case individually. Regions of elevated ensemble spread align with dynamically active areas of the magnetosphere, indicating that forecast uncertainty is largest where the system is intrinsically more variable.

Mid-range values of the standard deviation (0.1–0.2) are attributed to the magnetosheath region, where it is almost homogeneously distributed, with a slight increase near the bow shock. The plasma dynamics in the magnetosheath are strongly turbulent and, in the ion-kinetic regime, are largely governed by nonlinear interactions of electromagnetic waves driven by mirror and cyclotron instabilities. The behavior of the hybrid-Vlasov solver in this case is highly dependent on the numerical resolution \cite{dubart2020resolution}, and the resulting plasma state can be affected by numerical artifacts in velocity space that a fluid-like machine learning model is not aware of. Nevertheless, the model still reproduces wave activity with high accuracy, as evidenced by the correspondence of density striations in the magnetosheath observed in the left and central panels.

Peak values of the standard deviation ($\sim 0.3$) are localized at the magnetopause and magnetotail current sheets. The dark red regions in the right column of Figure~\ref{fig:rho} correspond to high-density structures, identified as plasmoids formed by magnetic reconnection (and flux transfer events at the dayside magnetopause, where applicable). The evolution of these structures is highly nonlinear, governed by ion acceleration processes that induce anisotropy and non-gyrotropy in the velocity distribution functions.

Because the model is trained only on fluid moments ($\rho, \mathbf{V}, P, T$), it lacks the velocity-space information required to capture reconnection onset, as well as the formation and evolution of plasmoids and flux transfer events. In particular, the standard deviation peaks in the cusp regions for Runs 2 and 3, where transient reconnection events drive anisotropic ion acceleration during secondary reconnection~\cite{jarvinen2018ion}.

The observed loss of physical accuracy in current-sheet regions is likely a direct consequence of the chosen state-vector representation. Two different VDFs can share the same lower-order moments at one instant yet evolve toward different moments at the next, so the moment-to-moment dynamics is not fully determined by the model's inputs. That is distinct from the classical closure problem of moment hierarchies. The missing information resides in higher-order velocity-space structure, such as non-thermal heat fluxes and pressure anisotropies, that regulates reconnection and associated instabilities. The present experiments cannot separate this insufficiency of the moment-only state from finite training data and finite model capacity. The elevated errors and ensemble spread localize where the moment-based description struggles, but do not by themselves establish the cause.

Appendix~\ref{app:results} provides additional results for other observables in the same format as Figure~\ref{fig:rho}. Figures~\ref{fig:bx} and \ref{fig:bz} show the magnetic field components $B_x$ and $B_z$ in the magnetotail and magnetopause current layers, respectively, where both exhibit uncertainty tied to their roles as reconnecting components. Figures~\ref{fig:ex}--\ref{fig:ez} display the electric field components $E_x$, $E_y$ and $E_z$ also associated with the reconnection process, while Figures~\ref{fig:vx} and \ref{fig:vz} present the corresponding reconnection outflow velocities $V_x$ (in the tail) and $V_z$ (at the magnetopause). All these diagnostics consistently indicate elevated standard deviation values in regions dominated by magnetic reconnection.

\subsection{Deterministic and probabilistic forecast evaluation}

\begin{figure}[ht]
\centering
\includegraphics[width=\textwidth]{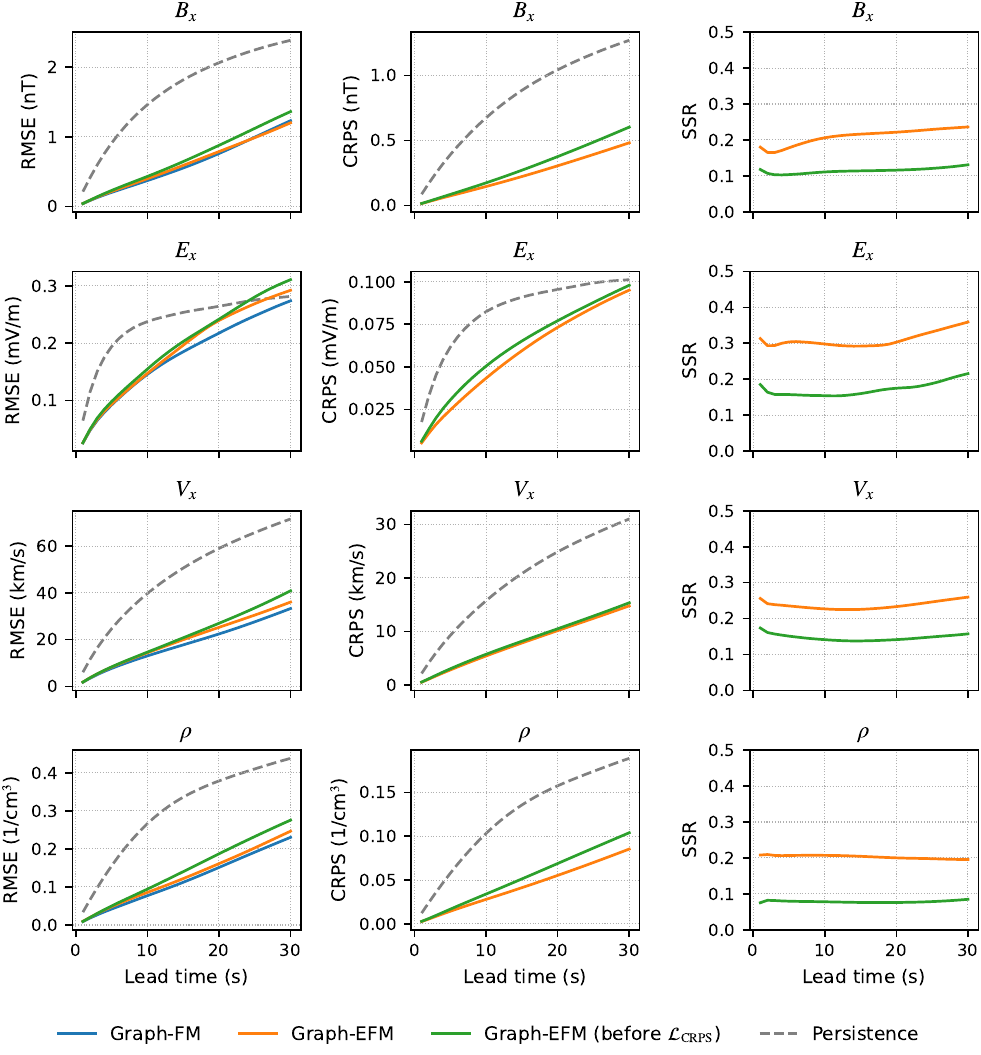}
\caption{Forecast performance on the test set indicated by RMSE, CRPS, and SSR for the deterministic (Graph-FM) and probabilistic (Graph-EFM) models across variables $B_x$, $E_x$, $V_x$, and $\rho$ for lead times 1--30\,s. Persistence is included as a baseline for RMSE and CRPS.}
\label{fig:metrics}
\end{figure}

The performance of all models is summarized in Figure~\ref{fig:metrics}, which shows the RMSE, CRPS, and SSR for lead times 1–30\,s. The metrics are calculated over the test set across the four runs such that forecasts are initialized at every timestep with an increment of 1\,s and rolled out for 30\,s. This yields 116 forecasts in total. Consecutive windows overlap within each run, making the samples correlated. The curves can be read as averages over the four test trajectories. Here we focus on forecasting performance for $B_x$, $E_x$, $V_x$, and $\rho$, with results for the remaining variables provided in Appendix~\ref{app:results} Figures~\ref{fig:metrics_2} and \ref{fig:metrics_3}.

The RMSE measures the accuracy of the deterministic forecast (or ensemble mean), while the CRPS evaluates both accuracy and calibration of the full predictive distribution. Overall, the RMSE results are similar for Graph-FM and Graph-EFM. For comparison, we also include Graph-EFM before CRPS fine-tuning. RMSE, CRPS and SSR are all clearly improved after fine-tuning with the CRPS loss. Some of this may be attributed to the model just undergoing more optimization steps with the other loss terms active as well. The RMSE and CRPS curves increase almost linearly with lead time because forecast errors accumulate gradually under autoregressive rollout.

As a reference, we include a persistence baseline, where the forecast is obtained by repeating the most recent observed state over the entire forecast horizon. For RMSE, this corresponds to the standard deterministic persistence forecast. For CRPS, which evaluates probabilistic predictions, we interpret persistence as a degenerate (Dirac) predictive distribution concentrated at the current state, i.e., a point mass. In this case, the CRPS reduces to the absolute error. This is also consistent with Eq.~\eqref{eq:crps}, since if all ensemble members take the same value, the ensemble spread term vanishes and the score reduces to the mean absolute error.

The emulators outperform the persistence baseline for both RMSE and CRPS across all lead times for most observables. However, for the out-of-plane magnetic field ($B_y$) and velocity ($V_y$), alongside the Hall electric field components ($E_x, E_z$), the errors tend to approach persistence more rapidly, and even exceed it at 30\,s for $E_x$. This phenomenon arises because these fields in particular are zero dominated across large portions of the simulation domain, which is hard for an autoregressive model to maintain. A similar trend is also observed when measuring the correlation with ground truth simulator states in Section~\ref{sec:correlation}, where the origin of the near-zero degeneracy in the 2D + 3V configuration and possible remedies are discussed in more detail.

We evaluate the probabilistic performance of Graph-EFM using the SSR, a standard metric for assessing the reliability of ensemble forecasts. The SSR compares the ensemble spread, the ensemble's internal estimate of its own uncertainty, to the actual forecast error measured by the RMSE. Values below unity correspond to \emph{underdispersive} ensembles, where the spread is too small relative to the forecast error, and values above unity indicate \emph{overdispersion}. Across all selected variables and lead times from 1\,s to 30\,s, we find SSR values in the range $0.2$--$0.3$. The SSR values are similar across variables and lead times, suggesting that Graph-EFM provides a stable but conservative estimate of forecast uncertainty across observables. Because the values are well below one, the model clearly underestimates its own uncertainty. This is a known characteristic of models trained primarily with a variational objective, and has also been observed in limited-area weather modeling \cite{oskarsson2024probabilistic}, a setting analogous to our magnetospheric setup in space. The SSR lines are all quite flat with respect to lead time indicating that ensemble spread and RMSE grow at similar rates, producing a temporally consistent uncertainty estimate.

Interpreting these results requires considering the sources of uncertainty represented by the model. In probabilistic forecasting, uncertainties are typically divided into \emph{epistemic} uncertainty, arising from limited data coverage or model capacity, and \emph{aleatoric} uncertainty, corresponding to irreducible variability in the underlying system. For our hybrid-Vlasov simulations, there is effectively no intrinsic aleatoric uncertainty, as the system is governed by a fully deterministic set of equations. Consequently, the uncertainty relevant for emulation is epistemic, reflecting the model’s incomplete knowledge of the dynamics. In contrast, machine learning weather forecasting models are trained on decades of reanalysis data that also contain aleatoric variability from observational noise injected into the simulation through data assimilation. In such settings, SSR values close to unity have been achieved~\cite{alet2025skillful}. Our lower SSR values indicate that the ensemble is under-dispersive, likely due to restricted training set and the lack of velocity-space information.

\subsection{Effect of magnetic divergence penalty}

\begin{figure}[ht]
\centering
\includegraphics[width=\textwidth]{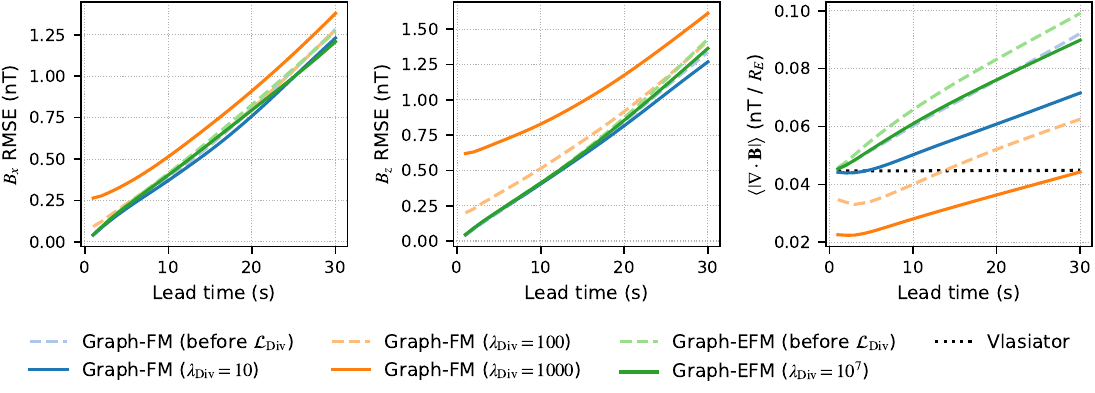}
\caption{
Effect of the magnetic divergence penalty. Shown are RMSE for $B_x$ and $B_z$ (left, center) and the mean absolute magnetic divergence over these fields (right) for Graph-FM and Graph-EFM models trained with different divergence-loss weights. The Vlasiator reference is included as a black dashed line. Moderate penalty weights reduce divergence without affecting RMSE, whereas excessively large weights can push $\nabla\cdot\mathbf{B}$ below the physical reference and increase the error of the corresponding magnetic field components.}
\label{fig:div_comparison}
\end{figure}

To examine the influence of the magnetic divergence penalty, we compare multiple variants of Graph-FM and Graph-EFM trained with different divergence-loss weights. Figure~\ref{fig:div_comparison} shows the RMSE for the magnetic field components $B_x$ and $B_z$ together with the mean absolute magnetic divergence $\langle |\nabla\cdot \mathbf{B}| \rangle$. The divergence penalty consistently reduces $\nabla\cdot \mathbf{B}$ across all models. For Graph-FM, weights of $\lambda_{\mathrm{Div}} = 10$ and $100$ yield progressively lower divergence, and $\lambda_{\mathrm{Div}} = 1000$ pushes $\nabla\cdot\mathbf{B}$ well below that of the Vlasiator reference. For Graph-EFM, a weight of $\lambda_{\mathrm{Div}} = 10^{7}$ provides an improvement in divergence freeness without degrading RMSE of the affected magnetic field components. Larger values could be explored for the probabilistic model, but we did not perform an exhaustive sweep due to the substantial computational cost associated with multi-step CRPS fine-tuning. Here divergence loss was applied during the final stage of training to isolate its impact, but it could also be applied from the beginning of training as it is not a computationally expensive metric to optimize.

The chosen values $\lambda_{\mathrm{Div}} = 10$ for Graph-FM and $\lambda_{\mathrm{Div}} = 10^{7}$ for Graph-EFM were selected so that the validation RMSE for $B_x$ and $B_z$ remained stable during training. The test-set curves confirm this: for these weights, the magnetic field RMSE is essentially unchanged relative to the baseline with no divergence loss. However, when the divergence penalty becomes too strong, as in the Graph-FM ablations with $\lambda_{\mathrm{Div}} = 1000$, the RMSE increases. This behavior arises from the competition between the data fidelity term and the divergence regularization. For moderate values of $\lambda_{\mathrm{Div}}$, the penalty acts as a physically motivated constraint that reduces spurious magnetic divergence without altering the predicted fields. However, when the weight becomes too large, the optimization increasingly prioritizes minimizing $\nabla\cdot\mathbf{B}$ over matching the target magnetic field values. As a result, the model adjusts the field components in a way that artificially suppresses divergence, which leads to a degradation in RMSE. In this regime the model effectively over-regularizes the solution and can even produce fields with lower numerical divergence than the Vlasiator reference when evaluated with the finite-difference discretization used here. That finite baseline is not an unphysical initialization. Vlasiator's field solver uses an upwind constrained transport scheme~\cite{londrillo2004divergence} that maintains $\nabla \cdot \mathbf{B} = 0$ to machine precision on its staggered, face-centered representation of the magnetic field~\cite{vonalfthan2014vlasiator}. The nonzero stationary value in Figure~\ref{fig:div_comparison} comes from applying second-order central differences to the cell-centered, volume-averaged fields of the reduced 1\,s output. That is not the representation in which the solver enforces the constraint, so the residual is a property of the diagnostic stencil. It does not grow in time and poses no issue for the simulation or for training. It does set the reference level for emulator-generated divergence. Driving the diagnostic below this baseline, as with $\lambda_{\mathrm{Div}} = 1000$, should not be read as improved physical fidelity.

\subsection{Ensemble size comparison}

\begin{figure}[ht]
\centering
\includegraphics[width=\textwidth]{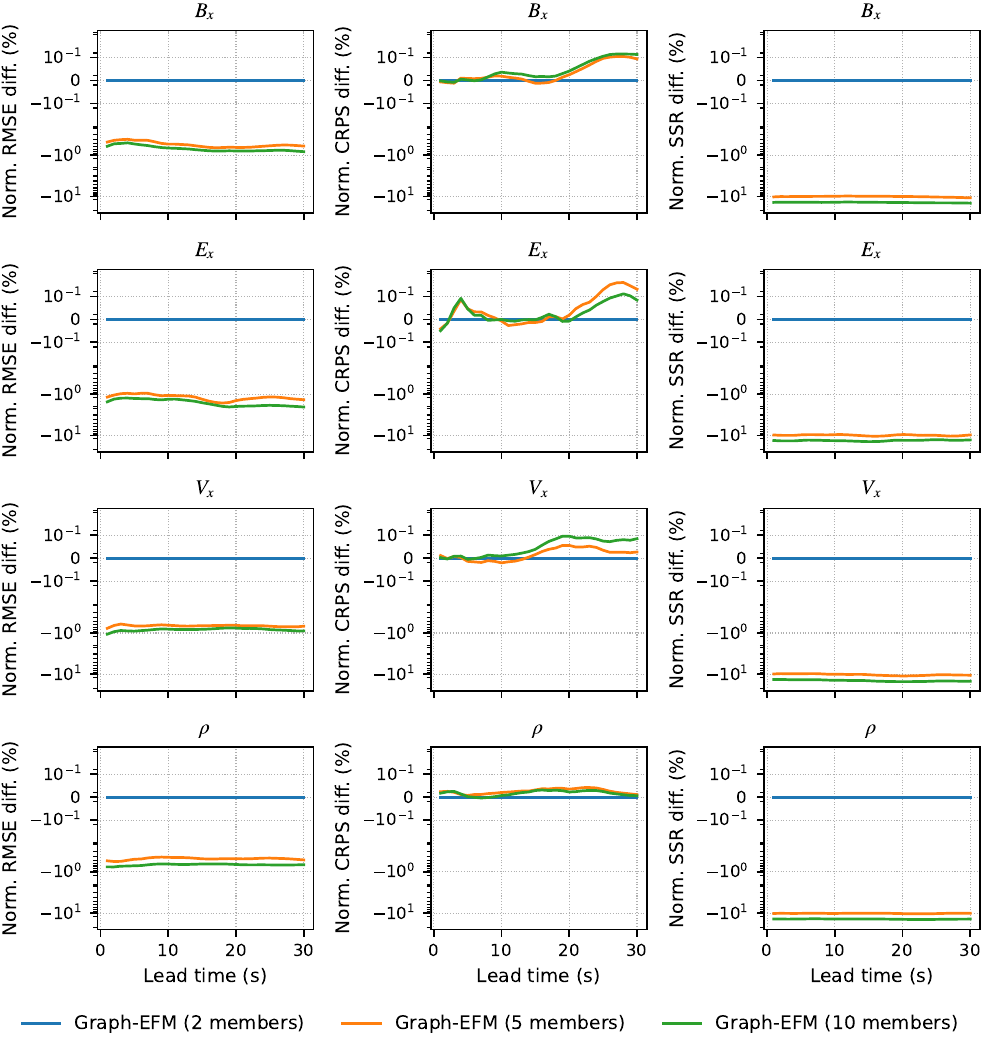}
\caption{
Normalized RMSE, CRPS, and SSR differences for Graph-EFM models using 2, 5, and 10 ensemble members. Values are reported relative to the 2-member ensemble (zero line). The y-axis has a symmetric logarithmic scale, that is linear between -0.2 and 0.2.}
\label{fig:ens_size}
\end{figure}

To assess how ensemble size affects predictive accuracy and uncertainty representation, we compare Graph-EFM models generating 2, 5, and 10 ensemble members. Figure~\ref{fig:ens_size} reports the normalized differences in RMSE, CRPS, and SSR relative to the 2-member ensemble, whose curves lie at zero by definition. For any metric $m$ and ensemble size $M$, we plot the quantity $\left(m_M/m_2 - 1\right)$, so negative values indicate a reduction relative to the 2-member baseline. Lower RMSE and CRPS differences reflect improved forecast accuracy. In contrast, higher values for normalized SSR differences are beneficial because the ensembles are underdispersive, with SSR values below one for all three models, and higher spread is desired for the ensembles to be considered well-calibrated.

Increasing the number of ensemble members consistently reduces RMSE. This behavior reflects variance reduction in the ensemble mean: with more samples drawn from the latent distribution, the ensemble mean becomes a more accurate estimator of the expected plasma state. In contrast, the CRPS remains essentially unchanged across ensemble sizes, as the 5- and 10-member ensembles fluctuate by less than 0.1\% relative to the 2-member ensemble. Because the CRPS evaluates the underlying predictive distribution rather than the number of samples drawn from it, its expected value is largely independent of ensemble size once the distribution is adequately represented.

The SSR metric, by comparison, decreases with increasing ensemble size. This could stem from a finite-sample effect, where occasional large pairwise differences increase spread, and larger ensembles provide a more stable estimate of the predictive variance while also reducing the RMSE of the ensemble mean.

\subsection{Autoregressive step size comparison}

\begin{figure}[ht]
\centering
\includegraphics[width=.95\textwidth]{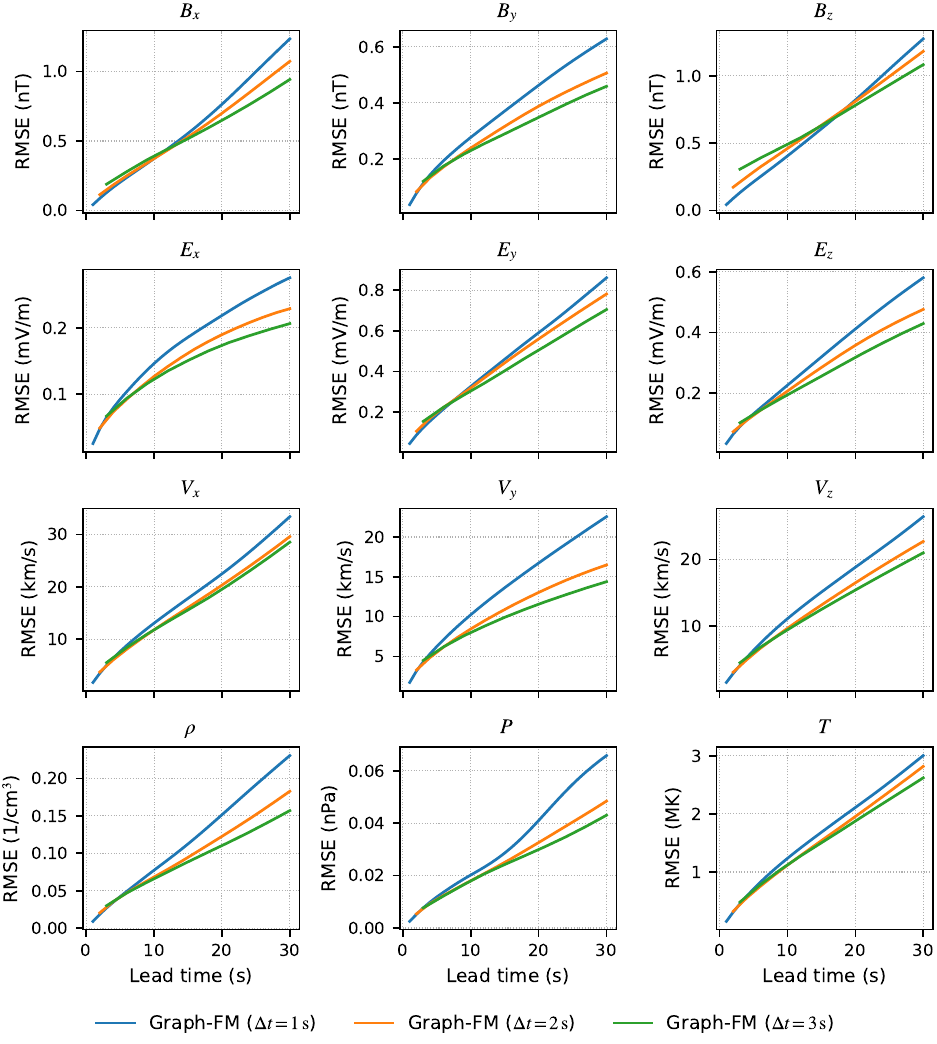}
\caption{RMSE as a function of lead time for Graph-FM models trained with autoregressive step sizes of $1$, $2$, and $3\,\mathrm{s}$. Larger step sizes reduce temporal error accumulation despite training on less data.}
\label{fig:step_size}
\end{figure}

To assess how the autoregressive stride affects forecast skill, we train Graph-FM models with $\Delta t = 1\,\mathrm{s}$, $2\,\mathrm{s}$, and $3\,\mathrm{s}$.  
The training data for the $2\,\mathrm{s}$ and $3\,\mathrm{s}$ models is obtained by temporal subsampling, reducing the number of available training snapshots by factors of two and three, respectively. Specifically, we apply a striding approach to the original $1\,\mathrm{s}$ simulation output by selecting every $k$-th snapshot (where $k=2,3$) to form the training sequences. To ensure a fair comparison, the number of optimization steps is held constant across models: the $2\,\mathrm{s}$ and $3\,\mathrm{s}$ variants are trained for proportionally more epochs so that each model performs an equal number of gradient updates.

Despite being exposed to fewer unique training samples, the $2\,\mathrm{s}$ and $3\,\mathrm{s}$ models accumulate less forecast error than the $1\,\mathrm{s}$ model across all variables in Figure~\ref{fig:step_size}, with the exception of a few early lead times. This improvement arises because longer autoregressive strides require fewer rollout steps to reach a given physical lead time, thereby reducing the compounding of one-step prediction errors. This is an inherent feature of neural emulators: unlike explicit physics solvers such as Vlasiator, where the time step is constrained by the fastest ion velocities and electromagnetic wave speeds relative to the grid spacing, neural models are statistical and not limited by such numerical stability requirements. A similar trend is observed in weather emulation, where models learn from 6-hourly mean aggregated reanalysis fields even though the underlying simulators run with much smaller time steps, and neural weather emulators continuously accumulate less error for 15-day forecasts as the step size is increased from 1\,h to 24\,h~\cite{bi2023accurate}.

\subsection{Power spectra of forecasted fields}

To further assess how well the models reproduce the spatial variability of the simulated plasma and field quantities, we compute the isotropic two-dimensional power spectra of each variable at several forecast lead times. For a given field $f(x,z)$, the power spectrum $P(k)$ is defined as the azimuthally averaged squared modulus of its two-dimensional Fourier transform,
\begin{equation}
P(k) = \left\langle \left| \hat{f}(k_x, k_z) \right|^2 \right\rangle_{(k_x,k_z):\sqrt{k_x^2 + k_z^2} = k},
\end{equation}
where $k_x$ and $k_z$ are the wavenumbers in the $x$- and $z$-directions, respectively, and $k = \sqrt{k_x^2 + k_z^2}$ denotes the radial wavenumber. The wavenumber $k$ has units of $R_E^{-1}$ and corresponds to spatial structures of characteristic scale $\lambda = \frac{2\pi}{k}$. Hence, small $k$ represents large-scale (global) structures, while large $k$ corresponds to finer, small-scale variations. All spectra are computed from 50\,s rollouts initialized at the start of the test segment for each of the four simulation runs, and then averaged across runs.

\begin{figure}[!ht]
\centering
\includegraphics[width=\textwidth]{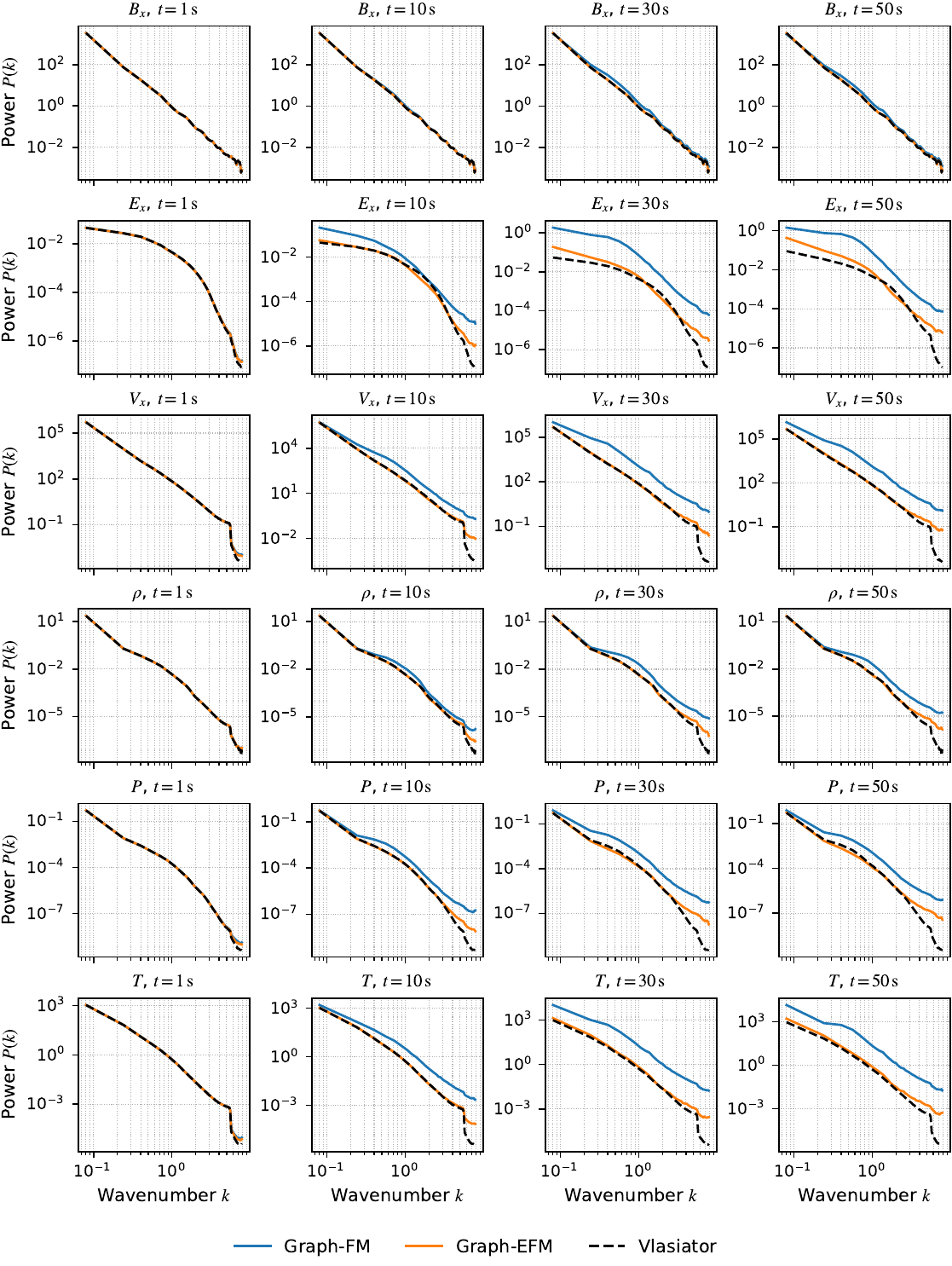}
\caption{
Spatial power spectra of $B_x$, $E_x$, $V_x$, $\rho$, $P$ and $T$. Each panel shows the isotropic power $P(k)$ versus wavenumber $k$, comparing Graph-FM and Graph-EFM forecasts with the Vlasiator ground truth at forecast lead times of $t = 1$, $10$, $30$, and $50\,\mathrm{s}$.}
\label{fig:power_spectra}
\end{figure}

The vertical axis in Figure~\ref{fig:power_spectra} shows the power $P(k)$, which measures the contribution of each spatial scale to the total variance of the field. Each subplot corresponds to one physical variable (rows) at a given forecast time (columns). The solid blue and orange curves indicate the spectra of the Graph-FM and Graph-EFM ensemble mean forecasts, respectively, while the black dashed curve shows the reference spectrum computed from the Vlasiator simulation. The models are evaluated at four lead times ($t = 1\,\mathrm{s}$, $10\,\mathrm{s}$, $30\,\mathrm{s}$, and $50\,\mathrm{s}$), allowing assessment of how well small- and large-scale energy distributions are preserved during the forecast. Appendix~\ref{app:results} Figure~\ref{fig:power_spectra_2} shows the power spectra of the remaining variables.

Both models reproduce the large-scale structure of the power spectra well, particularly for the magnetic field component $B_x$, whose spectrum closely matches the Vlasiator reference at all lead times. For the remaining variables ($E_x$, $V_x$, $\rho$, $P$, and $T$), differences emerge primarily at higher wavenumbers. Graph-EFM maintains spectra that remain close to the reference for most scales, with deviations confined to the very highest wavenumbers, indicating that it smooths away some fine-scale details over time, consistent with the spectral blurring commonly reported for machine learning forecast models at longer lead times~\cite{lam2023graphcast}. Graph-FM, on the other hand, shows elevated power across a wide band of high wavenumbers that stems from a buildup of small-scale noise or artifacts under autoregressive rollout.

An exception to this trend is the $B_x$ component in Figure~\ref{fig:power_spectra} and the $B_z$ component in Figure~\ref{fig:power_spectra_2}, whose spectra are reproduced with remarkable accuracy by both models at all lead times. This behavior can potentially be attributed to the intrinsic dipole magnetic field. In the inner magnetospheric region, the dipole field dominates the magnetic configuration and is orders of magnitude stronger than the perturbation field generated by magnetospheric dynamics, particularly close to the inner boundary, and this background field is comparatively static in time. As a result, $B_x$ and $B_z$ then inherit a strong, slowly varying large-scale structure that may stabilize their spectral content and makes it easier to reproduce accurately, even at longer lead times. 

\subsection{Correlation with simulation}
\label{sec:correlation}

Lastly, to assess forecast quality beyond aggregate error metrics, we analyze the pointwise correlation between predicted and true fields using prediction–versus–truth density plots. The analysis is based on the same 50\,s autoregressive rollouts initialized at the start of the test segment for each of the four simulation runs that were used in the spectral evaluation. For the results shown here, all grid points from the four forecasts are pooled for each physical variable and lead time.

\begin{figure}[!ht]
\centering
\includegraphics[width=.96\textwidth]{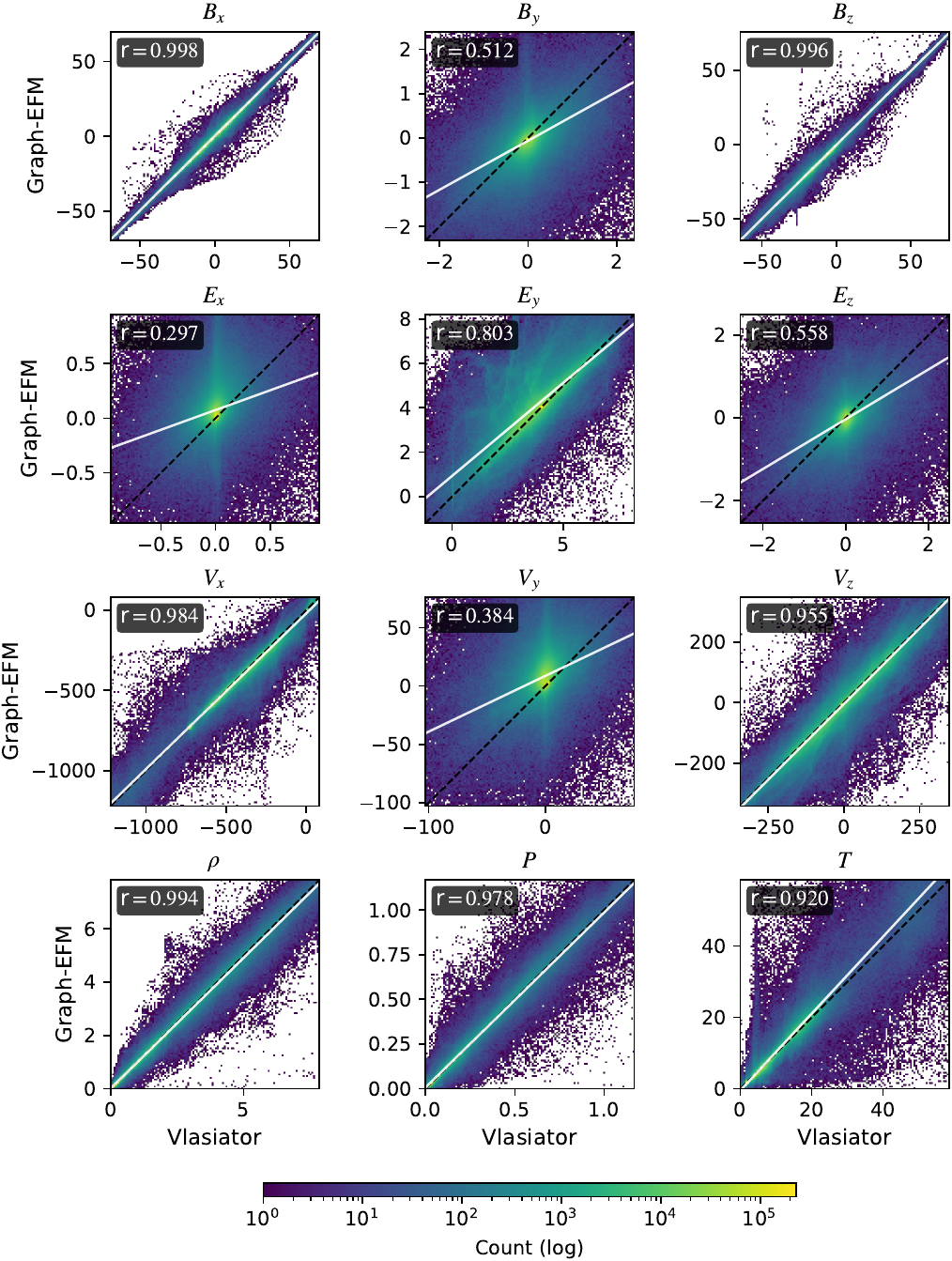}
\caption{
Two-dimensional density plots illustrating the relationship between predicted and ground truth values for Graph-EFM at a lead time of $t=30\,\mathrm{s}$ are shown for each physical variable. Each panel aggregates results from four forecasts. Colors represent logarithmic point density, the dashed line indicates the one-to-one relation, the solid line shows the least-squares fit, and the annotated Pearson coefficient summarizes linear correlation.
}
\label{fig:pred_vs_true_by_var_t30}
\end{figure}

Figure~\ref{fig:pred_vs_true_by_var_t30} shows the density plots for the Graph-EFM model at lead time $t=30\,\mathrm{s}$, displayed separately for each of the twelve physical variables. Each panel presents a two-dimensional histogram of predicted values against the corresponding Vlasiator reference values, with colors indicating point density on a logarithmic scale. The dashed black line denotes the ideal one-to-one relationship, the solid white line shows a least-squares fit, and the annotated Pearson correlation coefficient summarizes the linear association between forecast and ground truth.

Across most variables, the distributions remain concentrated along the one-to-one line, indicating that Graph-EFM preserves the dominant spatial structure of the simulated fields even at this extended lead time. As expected under autoregressive rollout, the distributions broaden relative to earlier lead times (see Appendix~\ref{app:results} Figures~\ref{fig:pred_vs_true_by_var_t10} and \ref{fig:pred_vs_true_by_var_t50} for lead times $t=10\,\mathrm{s}$ and $50\,\mathrm{s}$, respectively), reflecting the gradual accumulation of forecast uncertainty.

At longer lead times, a pronounced vertical band of high point density appears around zero true values in the prediction--truth histograms of Figure~\ref{fig:pred_vs_true_by_var_t30} for $B_y$, $E_x$, $E_z$, and $V_y$. This pattern arises because these field components are close to zero over large fractions of the simulation domain. The reason for this is that in the 2D + 3V hybrid-Vlasov configuration, spatial variations are restricted to the $x$--$z$ plane, i.e.\ $\partial/\partial y = 0$, which imposes an approximate symmetry in the out-of-plane direction. As a result, the dominant magnetospheric dynamics are primarily in-plane, producing strong $B_x$ and $B_z$, while the out-of-plane magnetic field $B_y$ is only generated through localized current systems and therefore remains small in most regions.

The same geometric constraint also shapes the electric field. In the hybrid model it is given by the generalized Ohm’s law:
\begin{equation}
\mathbf{E} = -\,\mathbf{V}\times\mathbf{B}
+ \frac{\mathbf{J}\times\mathbf{B}}{n e},
\end{equation}
where the first term represents the convective electric field and the second the Hall contribution. Because the symmetry implies no sustained force balance in the $y$ direction, the mean bulk flow satisfies $\langle V_y\rangle \approx 0$ over most of the domain, and the magnetic field is likewise predominantly in-plane with $B_y \approx 0$ except in localized regions. Under these conditions, the convective term reduces to:
\begin{equation}
-\,\mathbf{V}\times\mathbf{B}
=
\bigl(-(V_y B_z - V_z B_y),\; -(V_z B_x - V_x B_z),\; -(V_x B_y - V_y B_x)\bigr)
\approx (0,\;E_y,\;0),
\end{equation}
so that it contributes mainly to the out-of-plane electric field component $E_y$.

The in-plane components $E_x$ and $E_z$ therefore arise primarily from the Hall term $\mathbf{J}\times\mathbf{B}/(n e)$, which becomes significant in regions where the current is not aligned with the magnetic field, such as thin current sheets, reconnection sites, and boundary layers. These processes are spatially localized, explaining why $E_x$ and $E_z$ remain comparatively weak over most of the domain.

These physical motivations explain why a large fraction of grid points are zero for the variables $B_y$, $E_x$, $E_z$, and $V_y$, and even modest prediction variability around zero leads to a visually prominent vertical band of point density in the plots comparing prediction with ground truth (Figures~\ref{fig:pred_vs_true_by_var_t30}, \ref{fig:pred_vs_true_by_var_t10} and \ref{fig:pred_vs_true_by_var_t50}). The impact of this effect increases with forecast horizon: at lead time $t=10\,\mathrm{s}$ the distributions remain tightly aligned with the diagonal and Pearson correlations exceed 0.9 even for these more challenging variables, whereas by $t=30\,\mathrm{s}$ correlations drop to about 0.4--0.5 and the vertical band becomes more pronounced. At $t=50\,\mathrm{s}$, correlations further decrease to 0.1--0.25. The remaining observables exhibit consistently high correlations, many of them exceeding 0.9 at lead times of 30\,s and 50\,s.

The gradual buildup of unintended nonzero values is most clearly visible for $E_x$ in Figure~\ref{fig:ex}, particularly in the bow shock of Run~1. Given the strongly zero-peaked target distributions of $B_y$, $E_x$, $E_z$, and $V_y$, small unbiased prediction errors around zero can accumulate under autoregressive rollout. This can lead to the observed biases, also seen in the power spectra for the $E_x$ in Figure~\ref{fig:power_spectra} as a drift from the ground truth line, even at smaller wavenumbers.

Zero-inflated observables of this type are relatively uncommon in neural-emulator benchmarks. Possible treatments include predicting signed log-magnitudes of the affected components, mixture outputs with a point mass at zero, or a loss that emphasizes the sparse active regions. In this dataset the degeneracy of $B_y$, $E_x$, $E_z$, and $V_y$ follows from the 2D symmetry $\partial/\partial y = 0$. Three spatial dimensions would remove that global suppression. Autoregressive drift on near-zero targets can still appear in 3D wherever a component is locally suppressed, so those treatments remain relevant.

\section{Discussion}

In this work, we presented neural surrogates of global hybrid-Vlasov simulations with built-in uncertainty quantification. The emulation, however, takes place in a 2D plane and on a regular grid. Global hybrid-Vlasov models are in general computationally so demanding that mesh refinement becomes essential, especially when modeling the full three spatial dimensions, because the physics necessitates higher spatial resolution in critical regions, like the bow shock and the magnetotail reconnection site near Earth, while lower resolution suffices in less important areas such as inflow and outflow boundaries~\cite{ganse2023enabling}, resulting in irregular data structures. GNNs were deliberately chosen as the architecture here as they have been shown to work well also for irregular domains. They have been used for predicting both the mean flow~\cite{mousavi2025rigno} and the full distribution~\cite{valencia2025learning} of partial differential equation (PDE) systems. The graph construction also allows transfer across static resolutions. The simulation grid enters the model only through the grid-to-mesh and mesh-to-grid edge sets, so those edges can be replaced for a different grid and the trained processor weights fine-tuned without starting from scratch. Recently physics-motivated adaptive mesh refinement has also been implemented for Vlasiator~\cite{kotipalo2024physics}, where the refinement is controlled directly by physical conditions. Emulating simulations whose mesh connectivity changes over time remains an open challenge. Graph-based architectures are a natural fit for that setting.

The current surrogate operates within a narrow region of parameter space. The four Vlasiator simulations used for training are a sparse sampling of possible system states, and the assumption of steady solar wind conditions further restricts the model to a specific dynamical regime. The low spread–skill ratio (SSR $\approx$ 0.2–0.3) is consistent with that limited diversity: the ensembles remain underdispersive, and the surrogate interpolates among a small number of known trajectories rather than providing reliable uncertainty estimates for out-of-distribution dynamics.

The neural surrogate, trained on electromagnetic fields and lower order moments of the VDFs, is high in uncertainty in regions where kinetic effects dominate. Addressing this limitation requires incorporating higher-order kinetic information into the model. A natural progression would be to forecast the entire VDFs over time, rather than just their moments. Here one could take inspiration from previous work on emulating gyrokinetic simulations~\cite{paischer2025gyroswin}, where a hierarchical vision transformer was used for evolving a 5D distribution function over time. Another potential methodology for this is the development of machine learning driven closures for capturing heat flux~\cite{miloshevich2026electron}. Implementing such closures would allow the surrogate to account for energy transport and dissipation without the prohibitive cost of emulating the entire velocity distribution function for every grid cell. At present, our results serve to quantify exactly where these fluid-like approximations fail, providing a clear diagnostic baseline for the necessity of VDF-aware closures in global magnetospheric surrogates.

In terms of improving the probabilistic forecasts, alternative diffusion-based~\cite{price2025probabilistic} and CRPS-based~\cite{alet2025skillful} models have also demonstrated skillful ensemble weather forecasts, even with finetuning on few autoregressive steps. However, compared to diffusion models, Graph-EFM is a faster model, as a single forward pass per member is sufficient, whereas diffusion models require multiple passes due to their iterative refinement process. On the other hand, CRPS-based ensemble weather forecasting models trained by matching only marginal distributions through a CRPS loss function, also allow for sampling ensemble members in a single forward pass through the network~\cite{alet2025skillful}. Such an approach could provide an interesting alternative to the one presented here. The training schedule can further be made a bit simpler compared to the Graph-EFM variational framework~\cite{larsson2025crps}.

A further limitation of neural emulators is that they generally struggle with error accumulation on long rollouts, causing simulated trajectories to diverge. This happens because the model gradually operates out of distribution with each autoregressive step. The forecast horizons evaluated here, 30\,s for the aggregate metrics and 50\,s for the spectral and correlation diagnostics, are short compared to the hundreds of seconds over which magnetospheric events are typically studied with Vlasiator. The present results are therefore short-horizon proof-of-concept emulation, not validated forecasting of complete magnetospheric events. Extending those horizons will require more diverse training data and would benefit from dedicated rollout-stability techniques. To combat error accumulation, techniques based on diffusion models have been proposed to do iterative spectral refinement~\cite{lippe2023pde}, or transport a system moving into regions of low probability back to equilibrium~\cite{pedersen2025thermalizer}. Another promising diffusion-based method to tackle cumulative error for probabilistic forecasts involves training models to forecast a future state in a single step while ensuring the temporal consistency of the forecast~\cite{andrae2024continuous}, which can be used to provide continuous time ensemble forecasts.

Another central challenge for neural emulators is how to enforce physical consistency. In this study, we apply a soft constraint in the loss function to encourage divergence-freeness of the magnetic field. The divergence penalty acts as a local, kinematic regularizer that discourages violations of $ \nabla \cdot B=0$, but it does not anchor the time evolution of the system to the governing Maxwell-Vlasov equations. Enforcing constraints from Faraday’s law, Ohm’s law, and the momentum and energy transport equations might reduce error accumulation during autoregressive rollout. Recent advances have proposed alternative strategies for enforcing hard physical constraints directly within generative models, where conservation laws are formulated in their integral form and satisfied through probabilistic control of variance~\cite{hansen2023learning}. Alternatively, inference-time correction frameworks have been introduced to impose arbitrary nonlinear constraints on pretrained flow-based models by guiding their generative trajectories toward physics-consistent states~\cite{utkarsh2025physics}. Differentiable physics is a complementary option we have not explored here. Components of the governing equations can be implemented in an automatic-differentiation framework and embedded in the training loop, so that those components are solved exactly rather than only encouraged through penalties. Examples include learned corrections inside a differentiable fluid solver~\cite{kochkov2021machine}, solver-in-the-loop training~\cite{um2020solver}, learned turbulence models~\cite{list2022learned}, and learned hidden-variable closures in a differentiable plasma fluid solver~\cite{joglekar2023hidden}. In our setting, a differentiable projection enforcing $\nabla \cdot \mathbf{B} = 0$ inside the rollout, or a differentiable Maxwell update on the predicted fields, would turn the soft divergence penalty into a hard guarantee. That would require redesigning parts of the deterministic and probabilistic training pipelines.

Finally, a recent trend is also to attempt learning across systems to facilitate transfer of information between them. To facilitate such approaches, community initiatives are gathering diverse physics simulations to create large-scale datasets for machine learning~\cite{ohana2024well}. These can then be used to train foundation models, that learn dynamics from multiple physical systems~\cite{mccabe2024multiple}. Similar efforts have also taken place for weather~\cite{bodnar2025foundation}, solar activity~\cite{roy2025surya}, and astronomy~\cite{parker2025aion} to name a few examples. High-fidelity plasma data as presented here could be a valuable addition to these growing scientific datasets. Likewise, magnetospheric prediction models can \emph{potentially} be improved by finetuning foundation models, allowing them to leverage insights from various physical domains and generalize better.

\section{Conclusion}

In this work, we developed graph-based neural emulators to reproduce the short-term evolution of plasma and electromagnetic fields in global hybrid-Vlasov simulations. As a proof of concept within a restricted simulation family, we trained on a set of Vlasiator runs spanning a controlled range of particle number densities and Alfvén Mach numbers under steady upstream conditions. Our results show that a state vector limited to fluid moments effectively captures large-scale dynamics, but it is insufficient to fully resolve the kinetic processes that govern reconnection and instabilities. Future global surrogates should therefore incorporate higher-order moments or learned closures to move beyond large-scale emulation.

Both deterministic and probabilistic formulations were evaluated to establish a blueprint for surrogate-assisted space plasma research. The deterministic model provides fast, single-point next-step predictions, while the probabilistic model generates ensembles of future states whose spread supplies spatially structured relative uncertainty information. Such capabilities are intended to enable rapid parameter studies and high-throughput ensemble generation, tasks that are currently computationally prohibitive for first-principles kinetic solvers. CRPS-based fine-tuning improves CRPS, though the ensembles remain underdispersive. A soft divergence penalty further reduces magnetic field divergence without degrading predictive performance. Power spectral analysis confirms that the emulators capture the dominant large scale structures of the hybrid-Vlasov system, with small scale deviations consistent with autoregressive error accumulation. 

Another central contribution of this study is the release of a hybrid-Vlasov dataset comprising electromagnetic fields and plasma moments provided as easily accessible Zarr stores, which enables the broader machine learning community to work with plasma simulations that incorporate ion-kinetic effects at global scale. The open codebase for the forecasting methods is further compatible with developments in machine learning-based atmospheric weather prediction that can help translate advances in a very active field of research to the domain of space weather. Because the models are based on GNNs these efforts are also directly extensible to different geometries or irregular grids. This flexibility is needed when attempting to emulate magnetospheric plasma simulation with mesh refinement, which is heavily used for simulation in three spatial dimensions.

The results demonstrate that neural emulators can serve as fast uncertainty-aware approximations for computationally intensive kinetic models. A key limitation of the present study is the reliance on a 2D dataset with a fixed interplanetary magnetic field orientation. Consequently, the model's robustness across broader solar-wind variability remains to be established, as does its performance in a full 3D + 3V hybrid-Vlasov configuration. Extending these approaches to longer forecast horizons, three dimensional domains, a wider range of solar wind driving conditions, and the prediction of full VDFs represent promising directions for future research on machine learning-based hybrid-Vlasov surrogates.


%
%

\data{The source code used to train and evaluate the machine-learning models is openly available on GitHub~\cite{spacecast2025} under MIT open-source license. The training data was produced using Vlasiator version 5.3.1~\cite{vlasiator2024zenodo}, and is hosted in an open repository~\cite{vlasiator2025mldata} under Creative Commons Attribution 4.0 license.}

\ack{This work was funded by the Research Council of Finland grants 361901 and 361902 (FAISER). Vlasiator was developed with support from the European Research Council’s starting grant 200141 (QuESpace) and consolidator grant 682068 (PRESTISSIMO). D.H. acknowledges support from the Fulbright-KAUTE Foundation Award for conducting research at UC Santa Barbara. D.H. wishes to thank Joel Oskarsson and Erik Larsson for helpful discussions on probabilistic weather modeling. M.P. acknowledges Research Council of Finland grant 352846 (Centre of Excellence in Research of Sustainable Space). Computing resources were provided by the LUMI supercomputer, owned by the EuroHPC Joint Undertaking and hosted by CSC–IT Center for Science.}

\section*{Conflict of interest}
\noindent The authors declare there are no conflicts of interest for this manuscript.

\clearpage

\appendix

\section{Data details}
\label{app:data}

Table~\ref{tab:dataset} summarizes all physical variables included in the Vlasiator dataset, listing their notation, units, and the residual standard deviation of each variable for each of the four simulation runs. These residuals are computed as the standard deviation of the temporal differences after each variable has been normalized to unit variance, and provide a measure of the intrinsic variability of each variable. We use these residual standard deviations as weights in the loss function when training our machine learning models as a way to normalize by the magnitude of the dynamics when predicting the evolution of the system. The plasma variables correspond to standard moments of the ion velocity distribution function: the zeroth moment gives the ion number density $\rho$, defined as the integral of the ion distribution function over velocity space; the first moment yields the bulk flow velocity $\mathbf{V}$, representing the mean direction and speed of ion motion; and the second central moments appear as the scalar pressure $P$ and derived temperature $T$, which measure the thermal energy density and the intensity of random ion motions about the mean flow. In addition to the variables a few static fields comprising the coordinates $(x,z)$ and the radial distance $r$ from the origin are included as extra positional information.

\begin{table}[ht]
\centering
\caption{Summary of all variables in the Vlasiator dataset, including notation, unit and residual standard deviation for each simulation run.}
\label{tab:dataset}
\begin{tabularx}{\textwidth}{lccXXXX}
\toprule
Variable & Label & Unit & Run 1 $\sigma_\mathrm{Res}$ & Run 2 $\sigma_\mathrm{Res}$ & Run 3 $\sigma_\mathrm{Res}$ & Run 4 $\sigma_\mathrm{Res}$ \\
\midrule
Magnetic field $x$-component & $B_x$ & nT & 0.116 & 0.105 & 0.101 & 0.096 \\
Magnetic field $y$-component & $B_y$ & nT & 0.124 & 0.120 & 0.109 & 0.121 \\
Magnetic field $z$-component & $B_z$ & nT & 0.132 & 0.132 & 0.144 & 0.146 \\
Electric field $x$-component & $E_x$ & mV/m & 0.143 & 0.103 & 0.088 & 0.093 \\
Electric field $y$-component & $E_y$ & mV/m & 0.310 & 0.209 & 0.189 & 0.216 \\
Electric field $z$-component & $E_z$ & mV/m & 0.192 & 0.153 & 0.142 & 0.168 \\
Velocity field $x$-component & $V_x$ & km/s & 9.748 & 6.971 & 5.488 & 5.385 \\
Velocity field $y$-component & $V_y$ & km/s & 10.08 & 7.144 & 5.697 & 6.096 \\
Velocity field $z$-component & $V_z$ & km/s & 10.97 & 8.044 & 6.410 & 6.100 \\
Particle number density & $\rho$ & 1/cm$^{3}$ & 0.007 & 0.010 & 0.015 & 0.016 \\
Plasma pressure & $P$ & nPa & 0.003 & 0.004 & 0.005 & 0.005 \\
Plasma temperature & $T$ & MK & 1.214 & 0.812 & 0.720 & 0.631 \\
\bottomrule
\end{tabularx}
\end{table}

\section{Additional results}
\label{app:results}

\begin{figure}[t]
  \centering
  \includegraphics[width=\textwidth]{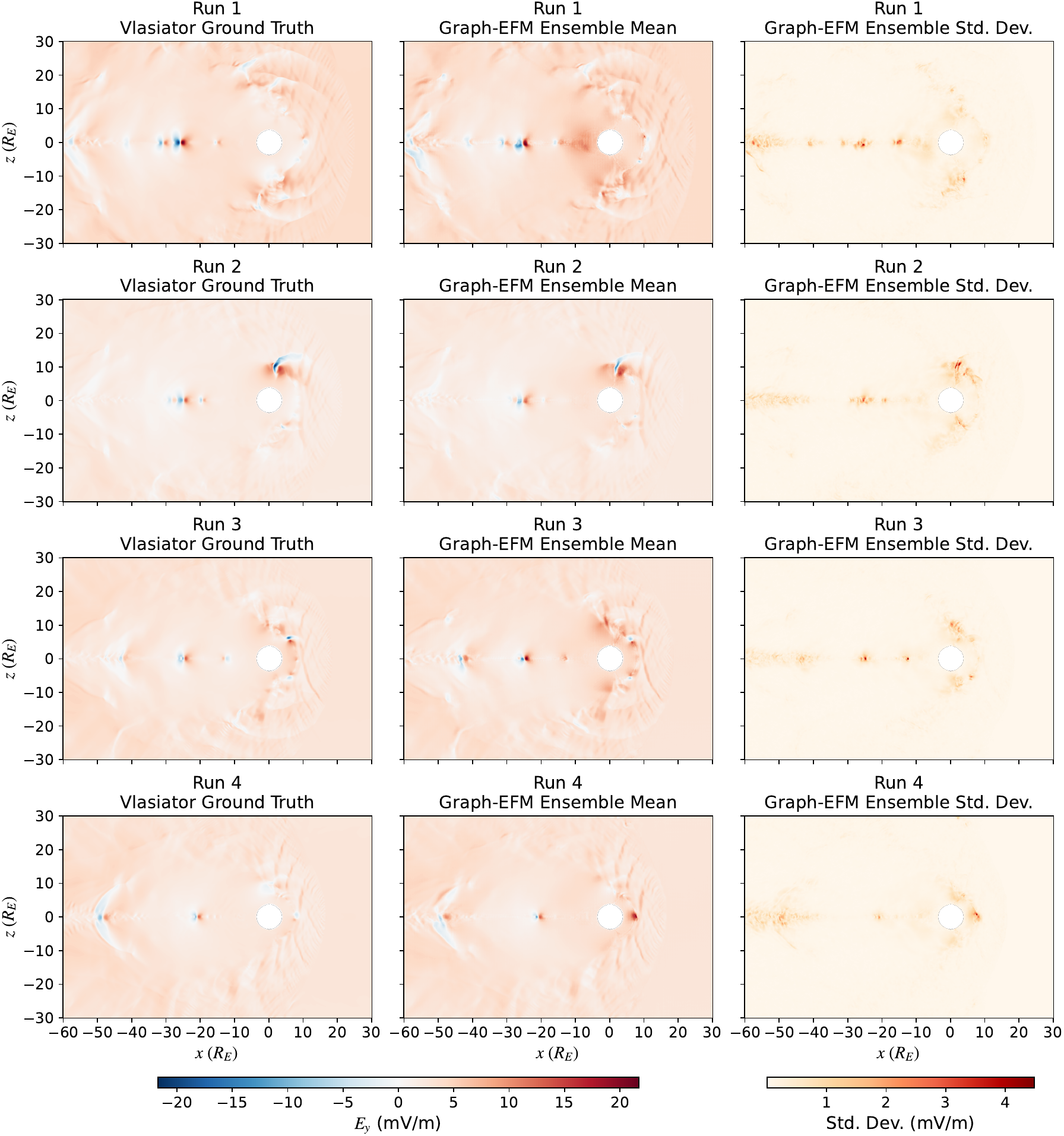}
  \caption{Example $E_y$ Vlasiator ground truth, Graph-EFM ensemble mean, and forecast uncertainty for each run at lead time $t=30\,\mathrm{s}$ for a forecast in the test set.}
  \label{fig:ey}
\end{figure}

Here we provide additional results and visualizations. We include example forecasts for the in-plane components of the electromagnetic and velocity fields, together with the out-of-plane electric field component $E_y$ shown in Figure~\ref{fig:ey}, which is of particular interest in this configuration because it is oriented along the direction where magnetic reconnection occurs. For example, in Run 4 there is a visible flux transfer event at the subsolar magnetopause, where Graph-EFM produces a strong reconnection electric field. We also provide forecast evaluation metrics, including RMSE, CRPS, and SSR, for all variables not shown in the main text. Similarly, the spatial power spectra for all remaining variables are included here that compare deterministic and probabilistic forecasts with the Vlasiator simulator. Finally, density plots comparing predicted values with the simulator are shown for lead times of $10\,\mathrm{s}$ and $50\,\mathrm{s}$, to give a better idea how these change over time.

\begin{figure}[h]
  \centering
  \includegraphics[width=\textwidth]{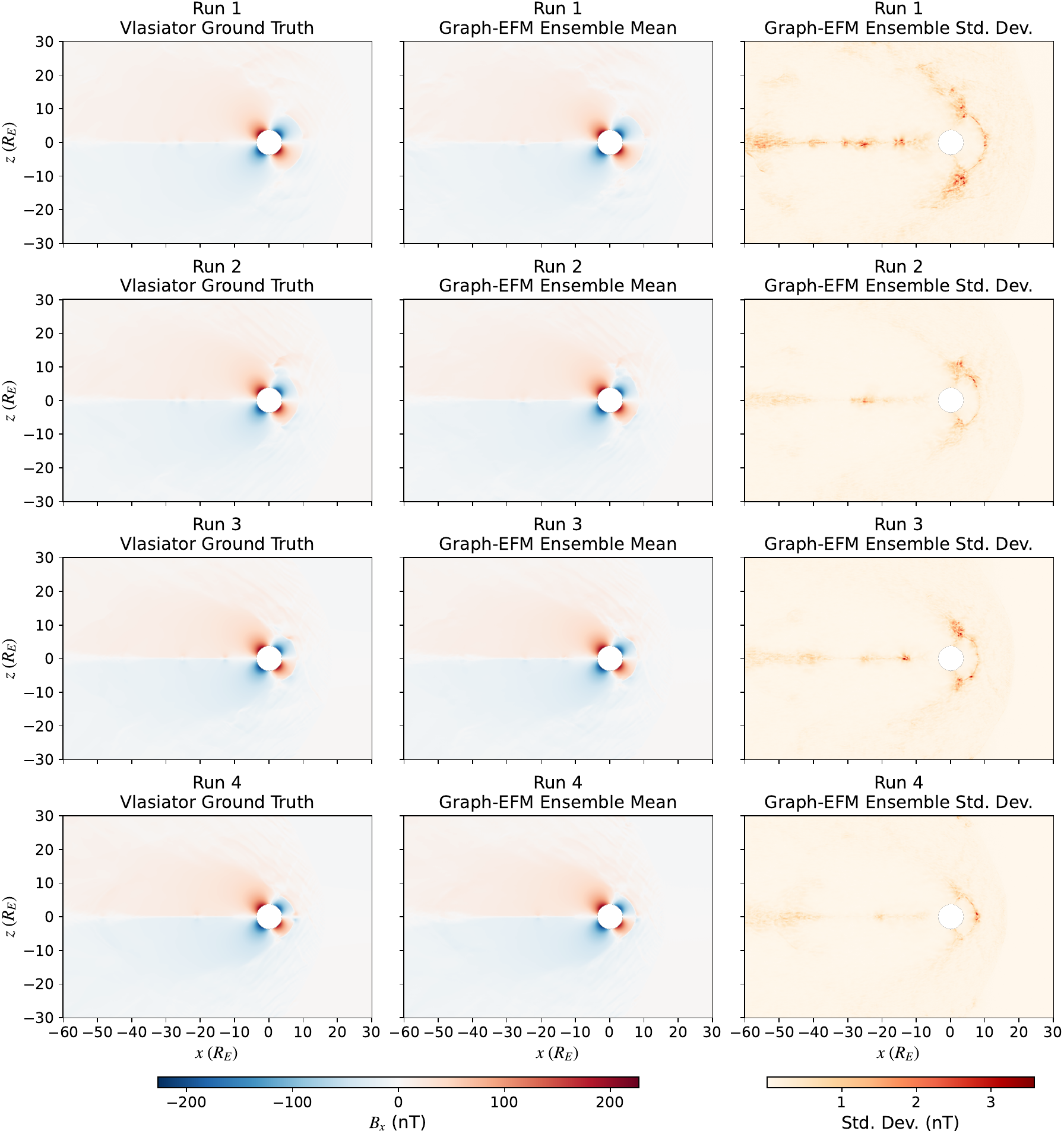}
  \caption{Example $B_x$ Vlasiator ground truth, Graph-EFM ensemble mean, and forecast uncertainty for each run at lead time $t=30\,\mathrm{s}$ for a forecast in the test set.}
  \label{fig:bx}
\end{figure}

\begin{figure}[t]
  \centering
  \includegraphics[width=\textwidth]{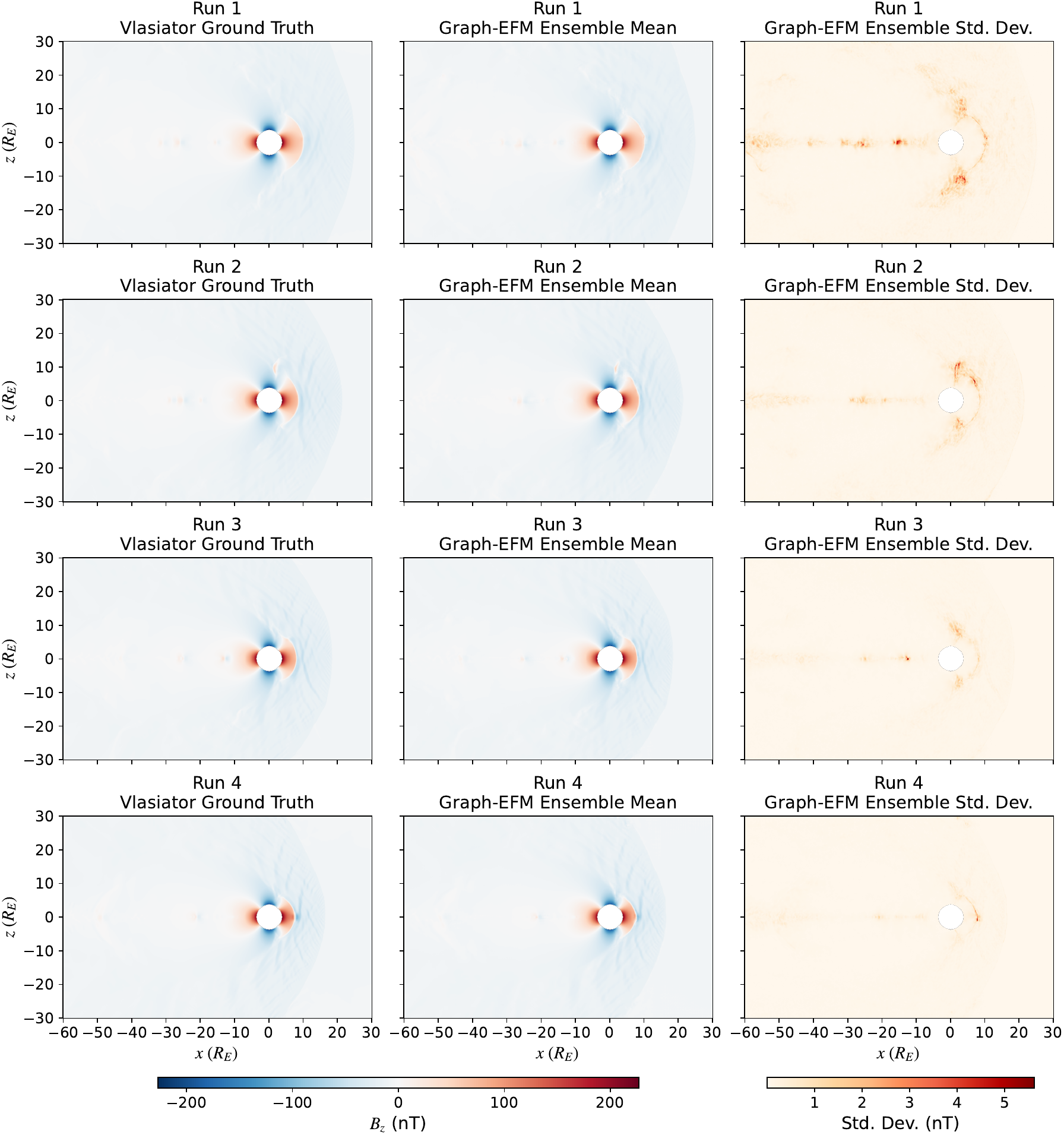}
  \caption{Example $B_z$ Vlasiator ground truth, Graph-EFM ensemble mean, and forecast uncertainty for each run at lead time $t=30\,\mathrm{s}$ for a forecast in the test set.}
  \label{fig:bz}
\end{figure}

\begin{figure}[t]
  \centering
  \includegraphics[width=\textwidth]{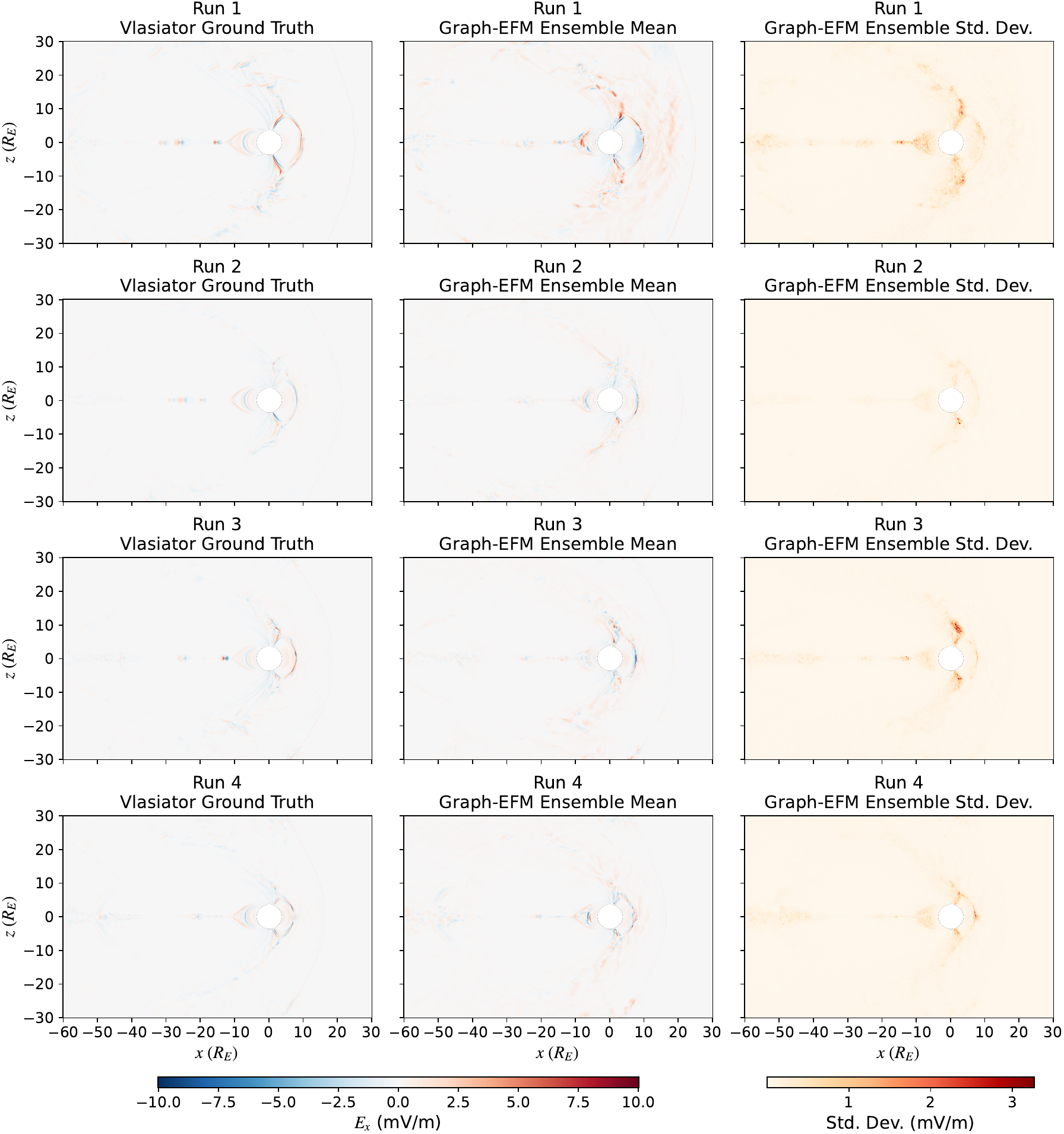}
  \caption{Example $E_x$ Vlasiator ground truth, Graph-EFM ensemble mean, and forecast uncertainty for each run at lead time $t=30\,\mathrm{s}$ for a forecast in the test set.}
  \label{fig:ex}
\end{figure}

\begin{figure}[t]
  \centering
  \includegraphics[width=\textwidth]{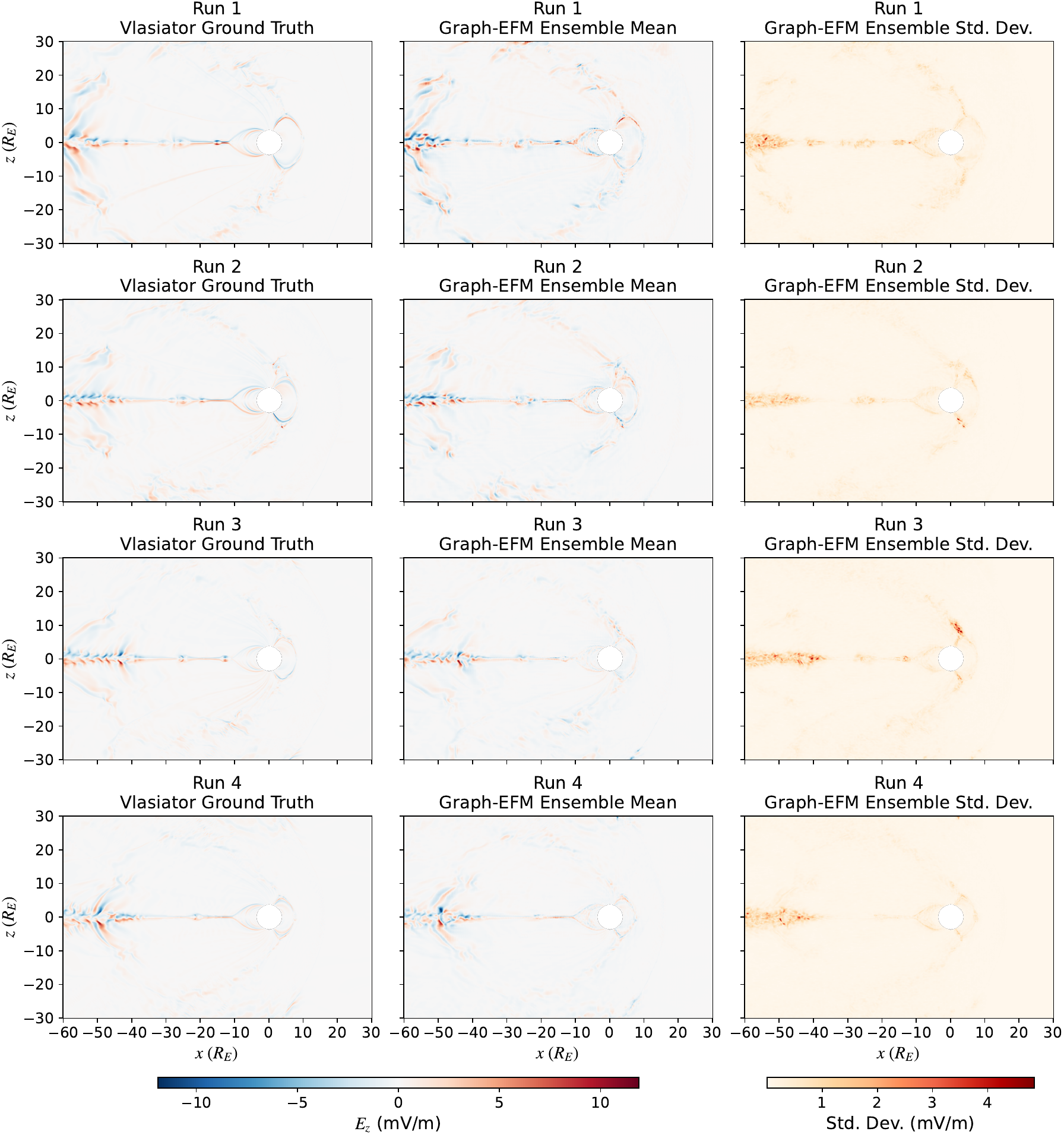}
  \caption{Example $E_z$ Vlasiator ground truth, Graph-EFM ensemble mean, and forecast uncertainty for each run at lead time $t=30\,\mathrm{s}$ for a forecast in the test set.}
  \label{fig:ez}
\end{figure}

\begin{figure}[t]
  \centering
  \includegraphics[width=\textwidth]{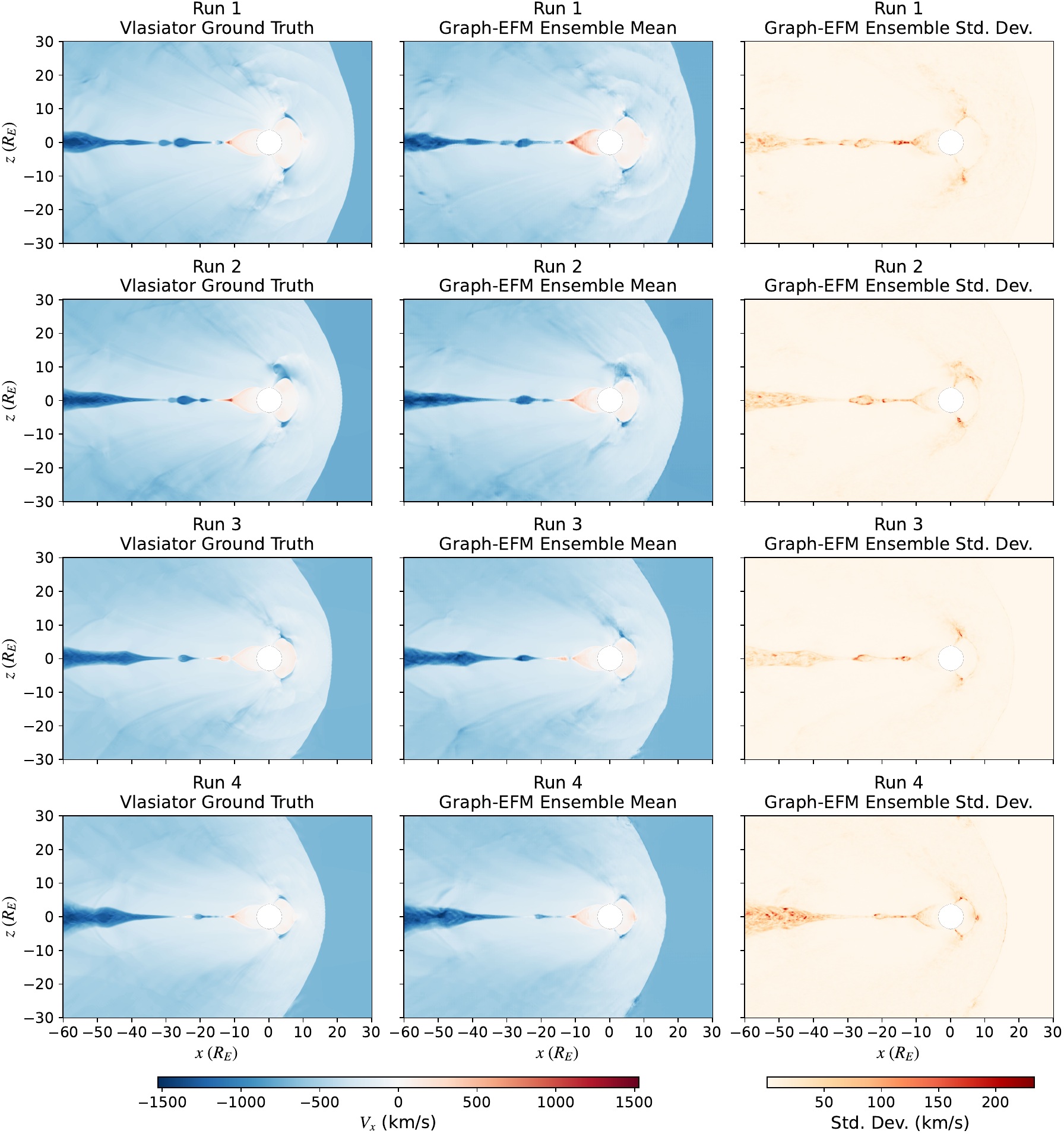}
  \caption{Example $V_x$ Vlasiator ground truth, Graph-EFM ensemble mean, and forecast uncertainty for each run at lead time $t=30\,\mathrm{s}$ for a forecast in the test set.}
  \label{fig:vx}
\end{figure}

\begin{figure}[t]
  \centering
  \includegraphics[width=\textwidth]{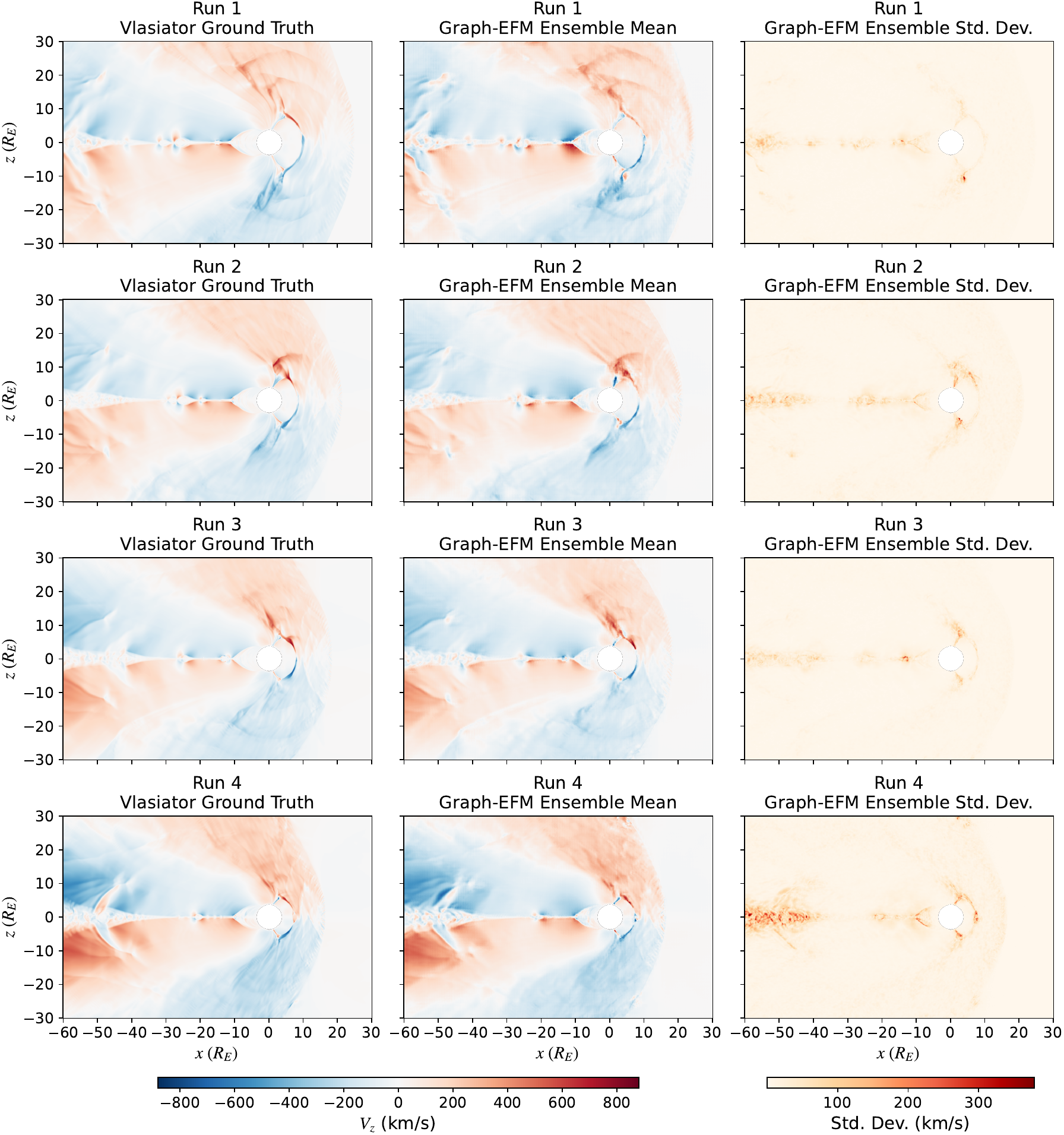}
  \caption{Example $V_z$ Vlasiator ground truth, Graph-EFM ensemble mean, and forecast uncertainty for each run at lead time $t=30\,\mathrm{s}$ for a forecast in the test set.}
  \label{fig:vz}
\end{figure}

\begin{figure}[ht]
\centering
\includegraphics[width=.97\textwidth]{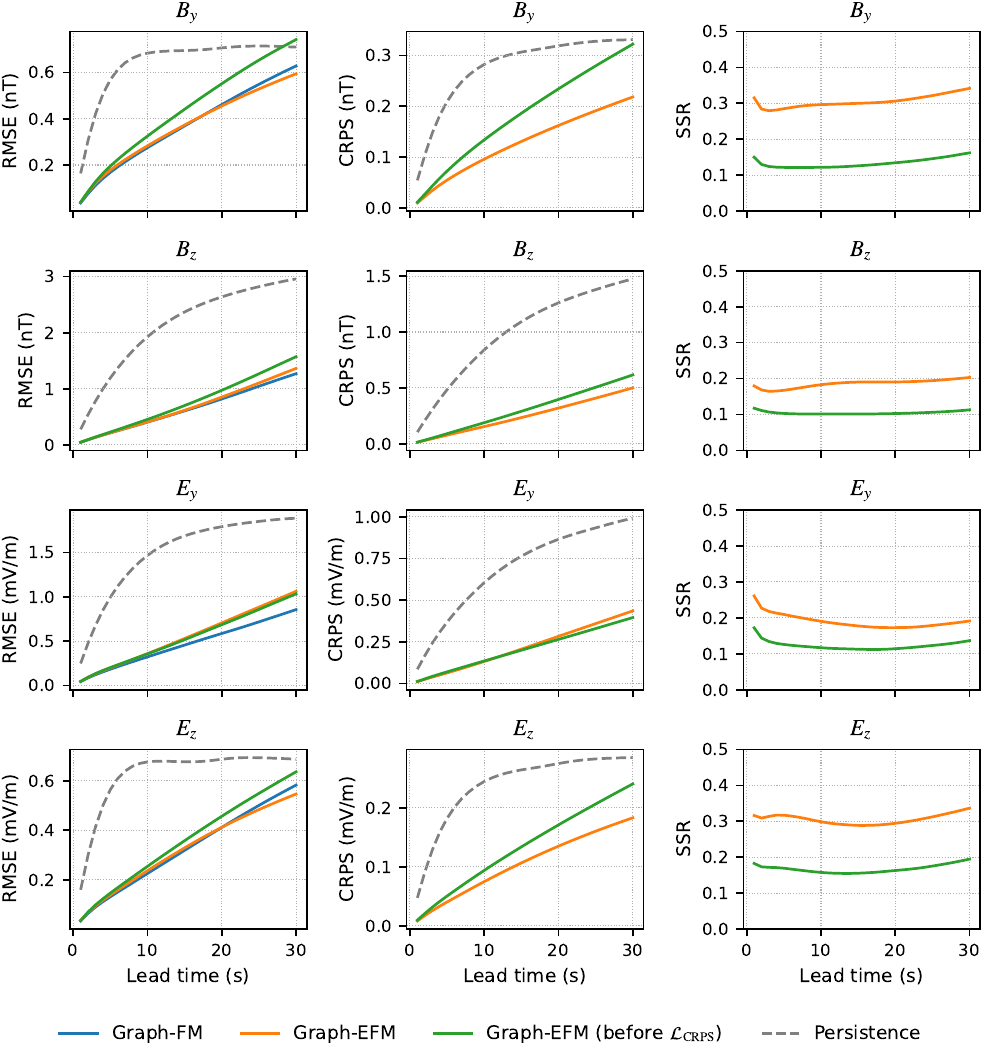}
\caption{Forecast performance on the test set indicated by RMSE, CRPS, and SSR for the deterministic (Graph-FM) and probabilistic (Graph-EFM) models across variables $B_y$, $B_z$, $E_y$, and $E_z$ for lead times 1--30\,s. Persistence is included as a baseline for RMSE and CRPS.}
\label{fig:metrics_2}
\end{figure}

\begin{figure}[ht]
\centering
\includegraphics[width=.97\textwidth]{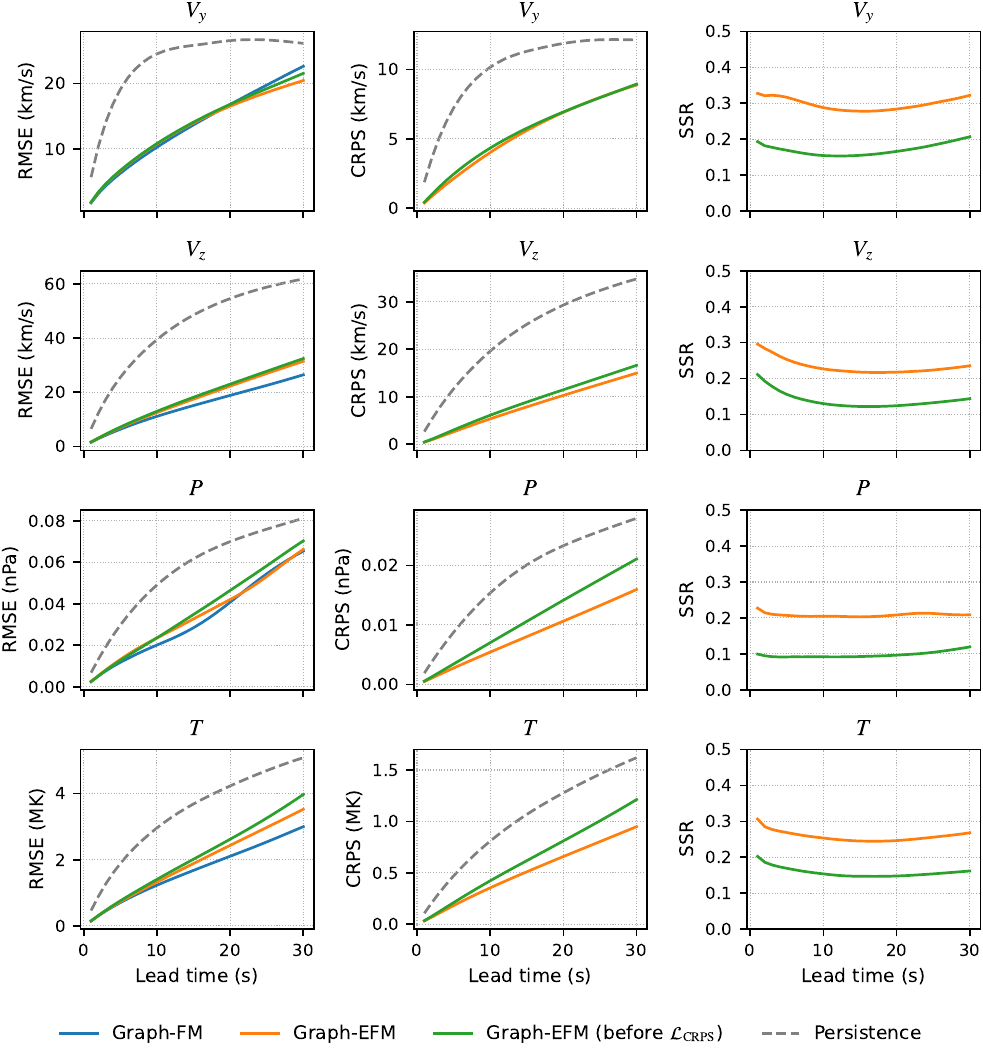}
\caption{Forecast performance on the test set indicated by RMSE, CRPS, and SSR for the deterministic (Graph-FM) and probabilistic (Graph-EFM) models across variables $V_y$, $V_z$, $P$, and $T$ for lead times 1--30\,s. Persistence is included as a baseline for RMSE and CRPS.}
\label{fig:metrics_3}
\end{figure}

\begin{figure}[ht]
\centering
\includegraphics[width=.97\textwidth]{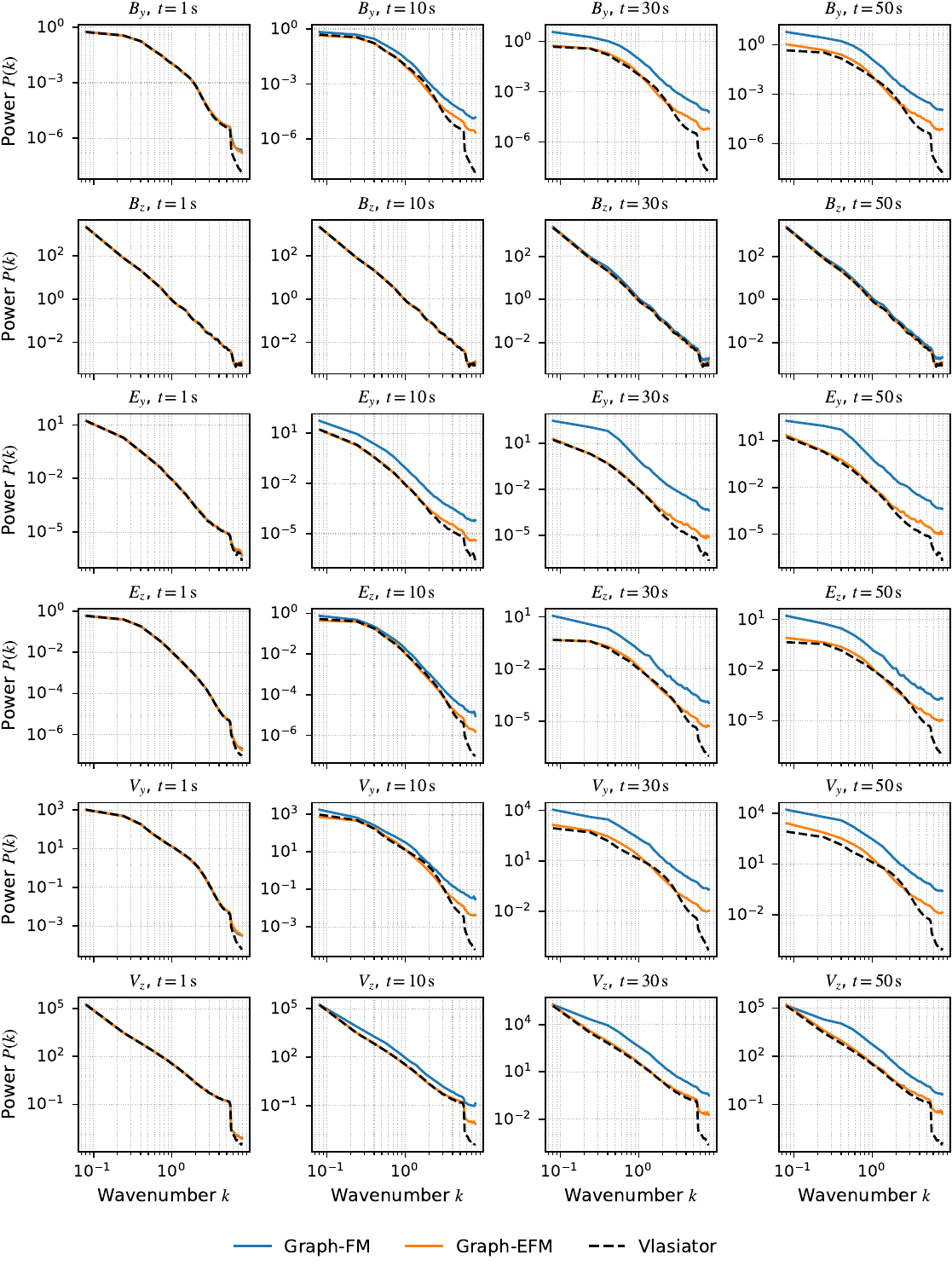}
\caption{
Spatial power spectra of $B_y$, $B_z$, $E_y$, $E_z$, $V_y$ and $V_z$. Each panel shows the isotropic power $P(k)$ versus wavenumber $k$, comparing Graph-FM and Graph-EFM forecasts with the Vlasiator ground truth at forecast lead times of $t = 1$, $10$, $30$, and $50\,\mathrm{s}$.}
\label{fig:power_spectra_2}
\end{figure}

\begin{figure}[!ht]
\centering
\includegraphics[width=.95\textwidth]{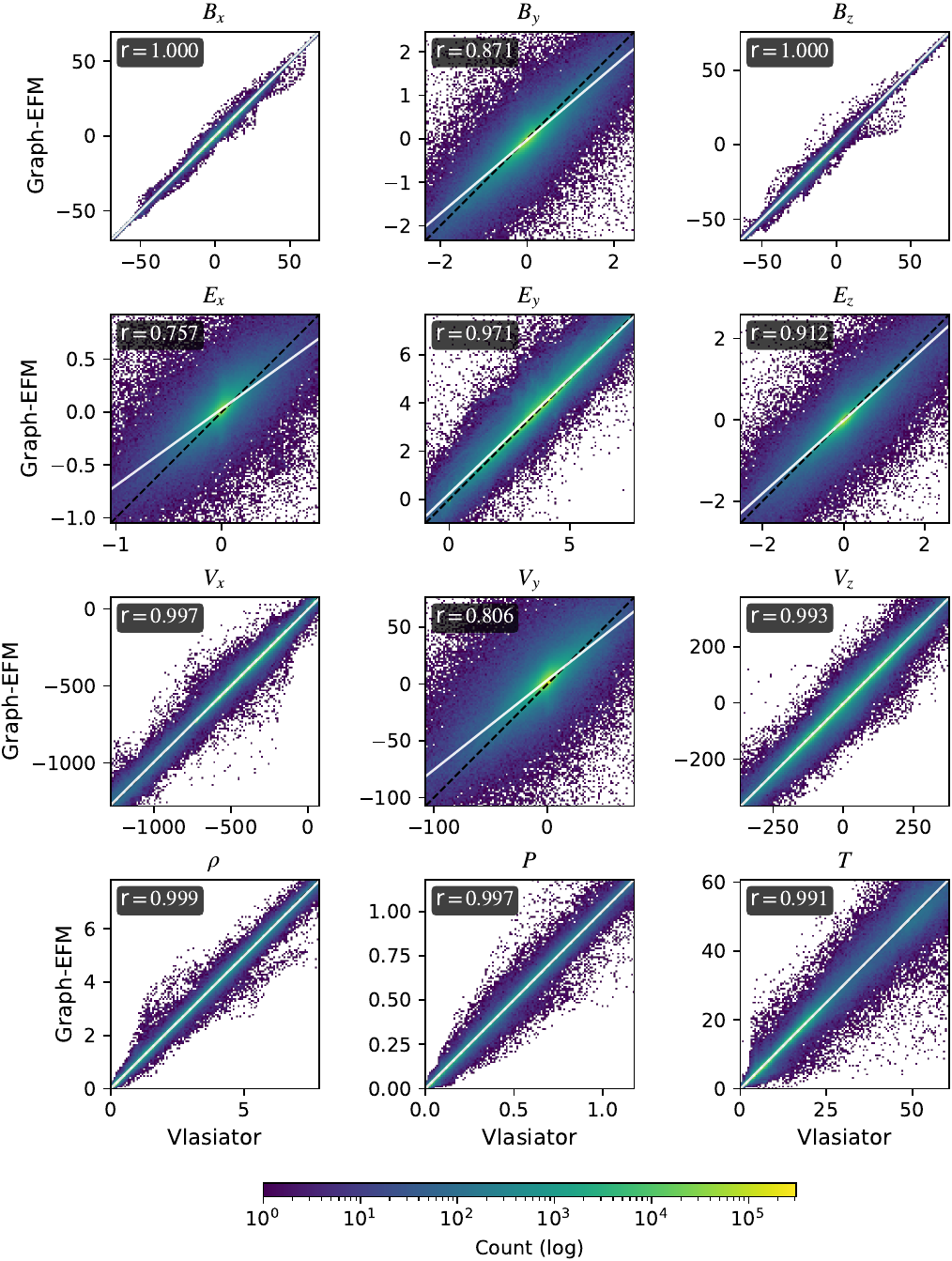}
\caption{
Two-dimensional density plots illustrating the relationship between predicted and ground truth for Graph-EFM at a lead time of $t=10\,\mathrm{s}$ are shown for each physical variable. Each panel aggregates results from four forecasts. Colors represent logarithmic point density, the dashed line indicates the one-to-one relation, the solid line shows the least-squares fit, and the annotated Pearson coefficient summarizes linear correlation.
}
\label{fig:pred_vs_true_by_var_t10}
\end{figure}

\begin{figure}[!ht]
\centering
\includegraphics[width=.95\textwidth]{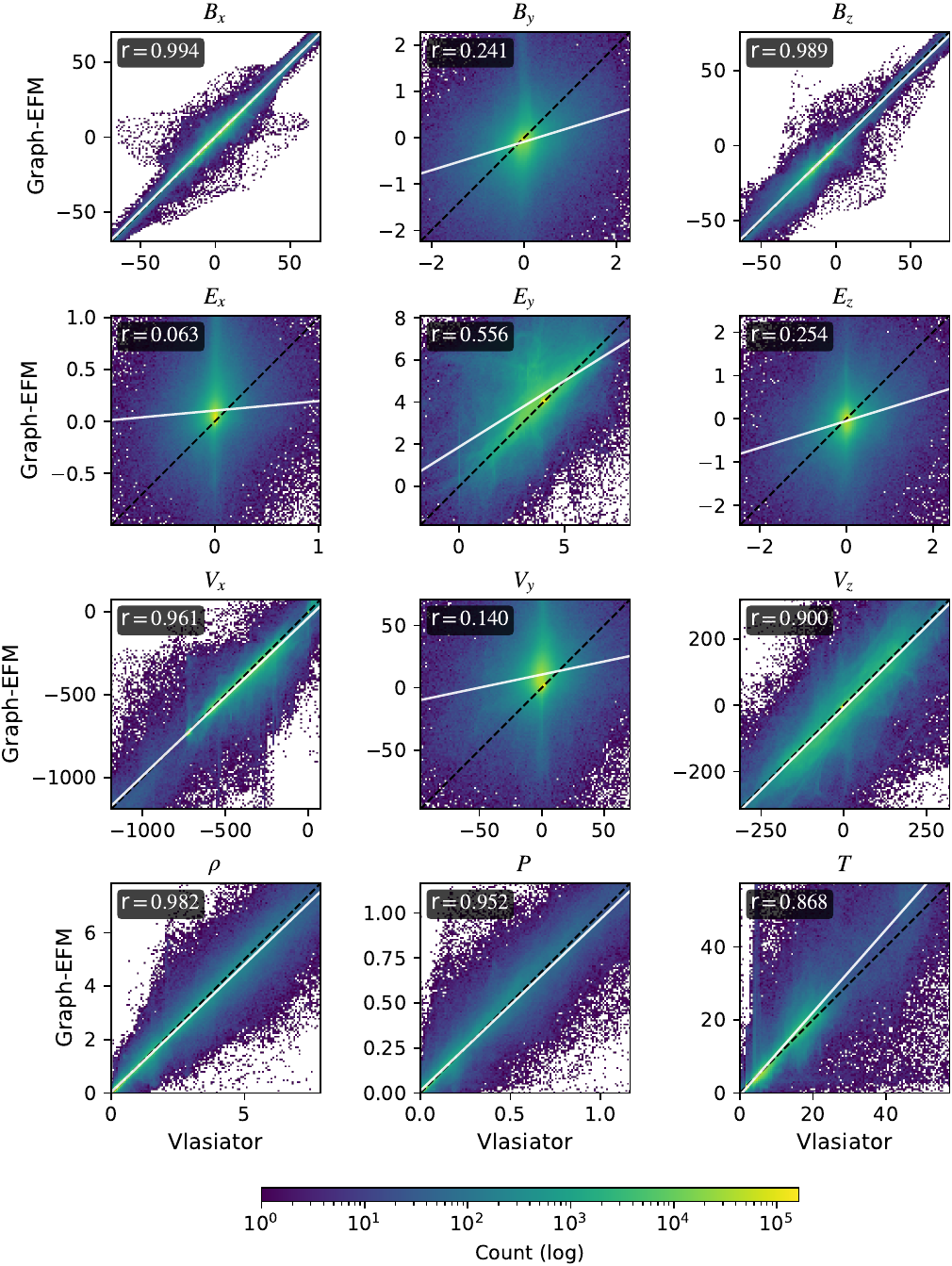}
\caption{
Two-dimensional density plots illustrating the relationship between predicted and ground truth for Graph-EFM at a lead time of $t=50\,\mathrm{s}$ are shown for each physical variable. Each panel aggregates results from four forecasts. Colors represent logarithmic point density, the dashed line indicates the one-to-one relation, the solid line shows the least-squares fit, and the annotated Pearson coefficient summarizes linear correlation.
}
\label{fig:pred_vs_true_by_var_t50}
\end{figure}

%
%

\clearpage

\bibliographystyle{iopart-num}
\bibliography{references}

\end{document}